\documentclass[twocolumn]{aastex62}
\usepackage{natbib}
\usepackage{amsmath,graphicx,hyperref}
\usepackage{natbib,threeparttable,rotating}
\usepackage{ulem}
\usepackage{multirow}
\newcommand{\etal}{et al.}

\newcommand{\hbeta}{H{$\beta$}}
\def\MgII{Mg\,{\sc ii}}
\def\CIV{C\,{\sc iv}}

\begin{document}

\title{The Sloan Digital Sky Survey Reverberation Mapping Project: The $M_{\rm BH}$-Host Relations at $0.2\lesssim z\lesssim 0.6$ from Reverberation Mapping and Hubble Space Telescope Imaging}

\author[0000-0003-1659-7035]{Jennifer I-Hsiu Li}
\affiliation{Department of Astronomy, University of Illinois at Urbana-Champaign, Urbana, IL 61801, USA}

\author[0000-0003-1659-7035]{Yue Shen}
\altaffiliation{Alfred P. Sloan Research Fellow}
\affiliation{Department of Astronomy, University of Illinois at Urbana-Champaign, Urbana, IL 61801, USA}
\affiliation{National Center for Supercomputing Applications, University of Illinois at Urbana-Champaign, Urbana, IL 61801, USA}


\author{Luis~C.~Ho}
\affiliation{Kavli Institute for Astronomy and Astrophysics, Peking University, Beijing 100871, China}
\affiliation{Department of Astronomy, School of Physics, Peking University, Beijing 100871, China}

\author{W.~N.~Brandt}
\affiliation{Department of Astronomy and Astrophysics, Eberly College of Science, The Pennsylvania State University, 525 Davey Laboratory, University Park, PA 16802, USA}
\affiliation{Institute for Gravitation \& the Cosmos, The Pennsylvania State University, University Park, PA 16802, USA}
\affiliation{Department of Physics, The Pennsylvania State University, University Park, PA 16802, USA}

\author{Elena~Dalla~Bont\`{a}}
\affiliation{Dipartimento di Fisica e Astronomia ``G. Galilei,'' Universit\`{a} di Padova, Vicolo dell'Osservatorio 3, I-35122 Padova, Italy}
\affiliation{INAF-Osservatorio Astronomico di Padova, Vicolo dell'Osservatorio 5 I-35122, Padova, Italy}

\author{G.~Fonseca Alvarez}
\affiliation{Department of Physics, University of Connecticut, 2152 Hillside Rd Unit 3046, Storrs, CT 06269, USA} 

\author{C.~J.~Grier}
\affiliation{Steward Observatory, The University of Arizona, 933 North Cherry Avenue, Tucson, AZ 85721, USA} 

\author{J.~V.~Hernandez Santisteban}
\affiliation{SUPA Physics and Astronomy, University of St. Andrews, Fife, KY16 9SS, Scotland, UK}

\author{Y.~Homayouni}
\affiliation{Department of Physics, University of Connecticut, 2152 Hillside Rd Unit 3046, Storrs, CT 06269, USA}

\author{Keith~Horne}
\affiliation{SUPA Physics and Astronomy, University of St. Andrews, Fife, KY16 9SS, Scotland, UK} 

\author{B.~M.~Peterson}
\affiliation{Department of Astronomy, The Ohio State University, 140 W 18th Avenue, Columbus, OH 43210, USA}
\affiliation{Center for Cosmology and AstroParticle Physics, The Ohio State University, 191 West Woodruff Avenue, Columbus, OH 43210, USA}
\affiliation{Space Telescope Science Institute, 3700 San Martin Drive, Baltimore, MD 21218, USA}

\author{D.~P.~Schneider}
\affiliation{Department of Astronomy and Astrophysics, Eberly College of Science, The Pennsylvania State University, 525 Davey Laboratory, University Park, PA 16802, USA}
\affiliation{Institute for Gravitation \& the Cosmos, The Pennsylvania State University, University Park, PA 16802, USA}

\author{Jonathan~R.~Trump}
\affiliation{Department of Physics, University of Connecticut, 2152 Hillside Rd Unit 3046, Storrs, CT 06269, USA}

\shorttitle{SDSS-RM: host galaxy imaging with HST}
\shortauthors{Li \etal}

\begin{abstract}
We present the results of a pilot Hubble Space Telescope (HST) imaging study of the host galaxies of ten quasars from the Sloan Digital Sky Survey Reverberation Mapping (SDSS-RM) project. Probing more than an order of magnitude in BH and stellar masses, our sample is the first statistical sample to study the BH-host correlations beyond $z>0.3$ with reliable BH masses from reverberation mapping rather than from single-epoch spectroscopy. We perform image decomposition in two HST bands (UVIS-F606W and IR-F110W) to measure host colors and estimate stellar masses using empirical relations between broad-band colors and the mass-to-light ratio. The stellar masses of our targets are mostly dominated by a bulge component. The BH masses and stellar masses of our sample broadly follow the same correlations found for local RM AGN and quiescent bulge-dominant galaxies, with no strong evidence of evolution in the ${M}_{\rm BH}-{M}_{\rm *,bulge}$ relation to $z\sim 0.6$. We further compare the host light fraction from HST imaging decomposition to that estimated from spectral decomposition. We found a good correlation between the host fractions derived with both methods. However, the host fraction derived from spectral decomposition is systematically smaller than that from imaging decomposition by $\sim 30\%$, indicating different systematics in both approaches. This study paves the way for upcoming more ambitious host galaxy studies of quasars with direct RM-based BH masses at high redshift.

\end{abstract}

\section{Introduction}

The observed local scaling relations between the masses of supermassive black holes (SMBHs) and their host-galaxy properties \citep[e.g.,][and references therein]{Magorrian_etal_1998,Gultekin_etal_2009,McConnell_Ma_2013,Kormendy_Ho_2013} are the cornerstone for the prevailing idea of the co-evolution between SMBHs and galaxies through some form of self-regulated black hole growth and feedback. A critical test of co-evolution scenarios and feedback models is to measure the evolution of the BH-host scaling relations beyond the nearby universe, and compare with theoretical work that implements various SMBH feeding and feedback recipes. In the past decade or so, large effort has been dedicated to measuring the host-galaxy stellar properties of distant (i.e., $z>0.3$) unobscured broad-line Active Galactic Nuclei (or quasars) using either imaging \citep[e.g.,][]{Treu_etal_2004,Treu_etal_2007,Peng_etal_2006a,Peng_etal_2006b,Jahnke_etal_2009,Merloni_etal_2010,Targett_etal_2012,Sun_etal_2015b} or spectroscopy \citep[e.g.,][]{ShenJ_etal_2008,Woo_etal_2006,Woo_etal_2008,Matsuoka_etal_2015,Shen_etal_2015b}. Combined with the BH mass measured using spectral methods \citep[e.g.,][]{Shen_2013} derived from local reverberation mapping results \citep[e.g.,][]{Peterson_2014}, these measurements were used to evaluate the correlations between SMBH mass and host properties beyond the local universe. This is currently the primary approach to measuring the evolution of the BH-host scaling relations. 

There are several challenges and caveats to this approach. First, host measurements are difficult due to the faintness of the galaxy and the contamination from the bright nucleus, requiring careful decomposition of the nuclear and host light. In the case of imaging, high spatial resolution is desired and sometimes necessary, and is often achieved with Hubble Space Telescope (HST). In terms of spectral decomposition (or decomposition of the broad-band spectral energy density), high S/N is required to separate the weak stellar continuum/absorption features from the bright quasar continuum. The second challenge of measuring BH-host properties at $z>0.3$
is that most of these distant samples have limited dynamic range in BH mass and only probe the high-mass end due to flux limit (for sufficient S/N), preventing the measurement of the BH-host correlations beyond simply inferring consistency or an ``offset'' from the local relations. There are a few recent exceptions where the dynamic range is more than an order of magnitude in BH mass \citep[e.g.,][]{Shen_etal_2015b,Matsuoka_etal_2015,Sexton_etal_2019}, allowing for the first time the determination of the slope and scatter of the correlations beyond the nearby universe. The third caveat, and perhaps the most significant one, is the large uncertainty of the BH mass estimates. So far all studies of the evolution of the BH-host scaling relations rely on BH masses estimated using the so-called ``single-epoch virial mass'' technique bootstrapped from local reverberation mapping results. These single-epoch masses have large systematic uncertainties (e.g., $\sim 0.4$ dex) that are fundamentally limited by the reverberation mapping sample \citep[see detailed discussions in, e.g.,][]{Shen_2013}. 

In addition to these inherent caveats, selection effects also play an important role in interpreting the observed ``evolution''. Neglecting selection effects, early studies based on small samples with a narrow dynamic range in mass often reported an excess of BH mass at fixed host properties from the local relations. Later more careful treatments of selection biases from the intrinsic scatter in the BH-host relation \citep[][]{Lauer_etal_2007}, BH mass uncertainties \citep[e.g.,][]{Shen_Kelly_2010},  or population biases \citep[e.g.,][]{Schulze_Wisotzki_2011}, combined with larger samples, have produced more cautious conclusions about the possible evolution of these scaling relations toward high redshift \citep[e.g.,][]{Schulze_Wisotzki_2014,Shen_etal_2015b,Sun_etal_2015b,Sexton_etal_2019,Ding_etal_2020}. These latest studies generally found that the results are consistent with non-evolving BH-host relations, at least to $z\sim 1$. Fully understanding these selection biases is difficult at this point, but future improvements in sample statistics and BH mass recipes will help reduce the statistical ambiguity in the interpretation of the observed evolution. 

In this work we lay the foundation for improving the constraints on the evolution of the $M_{\rm BH}-M_*$ relation, using a subset of 10 quasars from the Sloan Digital Sky Survey Reverberation Mapping \citep[SDSS-RM, ][]{Shen_etal_2015a} project for which we have acquired HST imaging data. 
The major advantage of our sample, compared with those used in most previous evolutionary studies, is that the BH mass estimates are based directly on reverberation mapping from a dedicated RM monitoring program, eliminating the systematic uncertainties associated with single-epoch BH masses. We use this sample as a pilot study to verify our methodology and to derive preliminary results on the evolution of the BH-host scaling relations using the SDSS-RM sample. 

This paper is organized as follows. In \S\ref{sec:data} we describe the sample and the HST data processing. We describe our imaging decomposition method in \S\ref{sec:analysis} and present the results in \S\ref{sec:results}. We discuss our results in \S\ref{sec:disc} and conclude in \S\ref{sec:con}. Throughout this paper we adopt a flat $\Lambda$CDM cosmology with $\Omega_M=0.3$ and $H_0=70\,{\rm km\,s^{-1}\,Mpc^{-1}}$. All host-galaxy measurements refer to the stellar population only.

\begin{table*}
\centering
\caption{Target Properties}
\label{tab:sample}
\begin{tabular}{ccccccccc}
\hline\hline
RMID & RA & DEC & z & ${i}_{psf}$ & ${\rm {L}_{5100, QSO}}$ & ${\rm {\sigma}_{*}}$ & log(${\rm {M}_{BH, SE}}$) & log(${\rm {M}_{BH, RM}}$) \\
 & [deg] & [deg] &  & [mag] & [erg/s] & [km/s] & [$\rm M_{\Sun}$] & [$\rm M_{\Sun}$] \\
\hline\hline
101 &213.0592 & 53.4296 & 0.4581 &18.84 & 44.4 & -- & 7.89$\pm$0.004 & ${7.26}_{-0.19}^{+0.17}$ \\
229 &212.5752 & 53.4937 & 0.4696 &20.27 & 43.6 & 130$\pm$8.7 & 8.00$\pm$0.07 & ${7.65}_{-0.20}^{+0.17}$ \\
272 &214.1071 & 53.9107 & 0.2628 &18.82 & 43.9 & -- & 7.82$\pm$0.02 & ${7.58}_{-0.21}^{+0.18}$ \\
320 &215.1605 & 53.4046 & 0.2647 &19.47 & 43.4 & 66.4$\pm$4.6 & 8.06$\pm$0.02 & ${7.67}_{-0.18}^{+0.18}$ \\
377 &215.1814 & 52.6032 & 0.3368 &19.77 & 43.4 & 115$\pm$4.6 & 7.90$\pm$0.03 & ${7.20}_{-0.16}^{+0.16}$ \\
457 &213.5714 & 51.9563 & 0.6037 &20.29 & 43.4 & 110$\pm$18 & 8.10$\pm$0.1 & ${8.03}_{-0.21}^{+0.18}$ \\
519 &214.3012 & 51.9460 & 0.5538 &21.54 & 43.2 & -- & 7.36$\pm$0.08 & ${8.99}_{-0.18}^{+0.17}$ \\
694 &214.2778 & 51.7278 & 0.5324 &19.62 & 44.2 & -- & 7.59$\pm$0.008 & ${6.70}_{-0.17}^{+0.35}$ \\
767 &214.2122 & 53.8658 & 0.5266 &20.23 & 43.9 & -- & 7.51$\pm$0.04 & $^*${${8.80}_{-0.17}^{+0.17}$} ({${8.26}_{-0.18}^{+0.20}$})\\
775 &211.9961 & 53.7999 & 0.1725 &17.91 & 43.5 & 130$\pm$2.6 & 7.93$\pm$0.008 & ${7.67}_{-0.24}^{+0.39}$ \\
\hline\hline
 \end{tabular}
 \tablecomments{$^*$The RM black hole mass of RM767 is calculated using the \MgII\, lag reported in \cite{Homayouni_etal_2020} and \citet[][value in brackets]{Shen_etal_2016a} and the broad \MgII\, FWHM measured from the mean spectrum from \cite{Shen_etal_2019b}. All other RM black hole masses are based on \hbeta\, lags from \cite{Grier_etal_2017}. The host stellar velocity dispersion $\sigma_{*}$ and single-epoch mass uncertainties are $1\sigma$ measurement errors only, while the RM mass uncertainties also include 0.16\,dex systematic uncertainty following \citet{Grier_etal_2017}. }
\end{table*}

\section{Data}\label{sec:data}
\subsection{SDSS-RM and Sample Selection}
The Sloan Digital Sky Survey Reverberation Mapping (SDSS-RM) project \citep{Shen_etal_2015a} has simultaneously monitored a uniform, flux-limited sample of 849 quasars in a 7 ${\rm deg}^{2}$ field since 2014 with both imaging and spectroscopy. The primary goal of SDSS-RM is to measure direct, RM-based BH masses for a uniform quasar sample that covers a broad luminosity and redshift range. As of June 2020, RM BH masses have been successfully measured for $\sim 150$ SDSS-RM quasars using multiple broad emission lines, including 18 with H$\alpha$ \citep{Grier_etal_2017}, 44 with H$\beta$ \citep{Shen_etal_2016a, Grier_etal_2017}, 57 with \MgII\ \citep{Shen_etal_2016a, Homayouni_etal_2020}, and 48 with \CIV\ \citep{Grier_etal_2019}.

Ten quasars with significant lag detections from the first year of  monitoring \citep{Shen_etal_2015a} were chosen for a pilot study of their host galaxies using HST imaging. The RM time lags and BH masses of these ten quasars are presented in \cite{Shen_etal_2016a}, \cite{Grier_etal_2017} and \cite{Homayouni_etal_2020}, and the host galaxy properties derived from spectral analysis are presented in \citet{Shen_etal_2015b} and \citet{Matsuoka_etal_2015}. These quasars spread over a factor of ten in luminosity within a redshift range of $0.2\lesssim z\lesssim 0.6$ (with $\left<z\right>=0.4$). Table \ref{tab:sample} summarizes the physical properties of the ten targets. 

\subsection{HST Imaging}

The ten quasars were observed with the Wide Field Camera 3 (WFC3) UVIS F606W filter and IR F110W filter in Cycle 23 (GO-14109; PI: Shen). 
To improve the point-spread-function (PSF) sampling, we used a basic 3-point dithering pattern for the F606W observations and a 4-point dithering pattern for the F110W observations. Multiple short exposures were used for the F606W observations to avoid saturation of the central point source. For IR F110W, we use the multi-step readout sequence (STEP) to correct for central-pixel saturation and to improve the dynamic range in the image. Two orbits were dedicated to each target, one for each filter.

To reliably subtract the central quasar light in the image, we construct PSF models by dedicating one orbit to observing the white dwarf EGGR-26 using the same filters and dithering patterns as our science observations. We group the observations within a 7-day window to minimize effects from optics changes of the instrument that may slightly change the PSF. Observations of seven targets (RM272, RM320, RM377, RM457, RM519, RM694, RM775) and the white dwarf were carried out between January 8$^{\rm th}$ and 17$^{\rm th}$ 2017. Initial visits for the remaining three targets (RM101, RM229, and RM767) failed and were repeated between March 6$^{\rm th}$ and 9$^{\rm th}$ 2017. There is no significant change in the quasar PSF of the later repeated observations for the remaining three targets, suggesting that the PSF is stable within the extended period of our observations.

We followed the standard HST pipeline procedures to reduce and calibrate these data with the best reference files provided by the HST Calibration Reference Data System (CRDS). The individual exposures are geometrically-corrected and dither-combined with {\it astrodrizzle}. We adjust the final pixel size ({\it final\_scale}) and pixel fraction ({\it final\_pixfrac}) following the {\it astrodrizzle} handbook to optimize the resolution of the drizzled images and to create a narrower, sharper PSF. The final image samplings are chosen to be 0.033$''$/pixel for the F606W images and 0.066$''$/pixel for the F110W images, which correspond to $\sim$0.18 and $\sim0.35\,$ kpc at $z=0.4$. Since the detector counts are conserved during the drizzling procedure, the chosen image sampling does not affect the photometry measurements. 

\begin{figure*}
\centering
\includegraphics[width=0.9\textwidth]{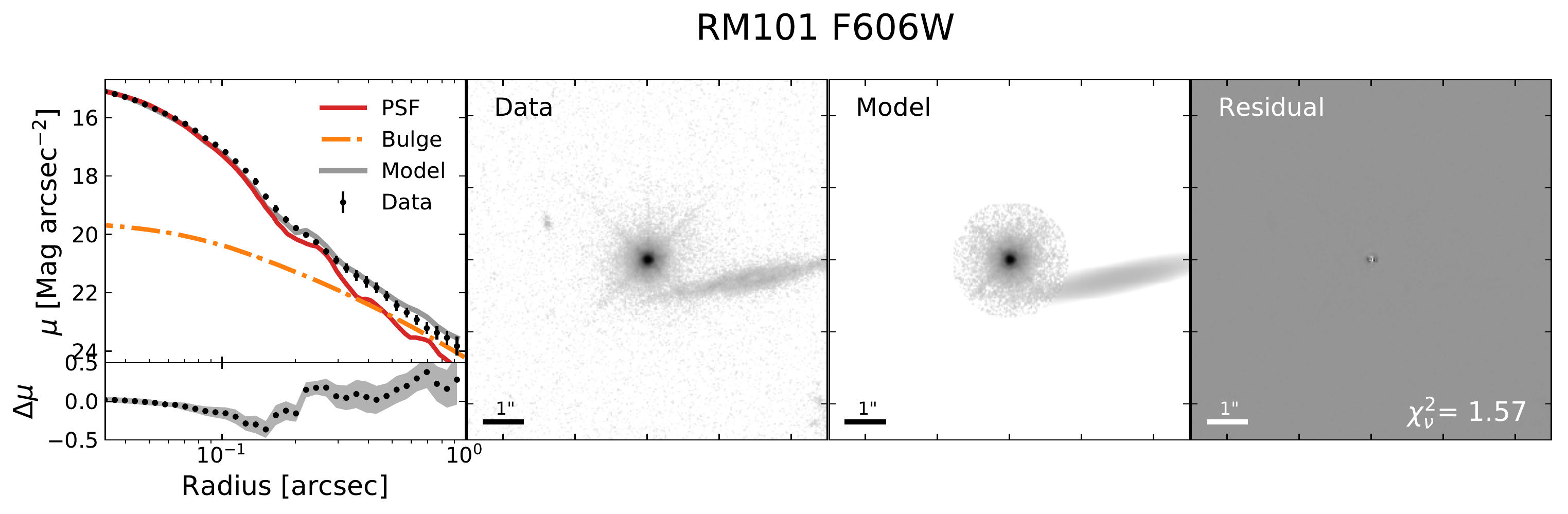}\\
\includegraphics[width=0.9\textwidth]{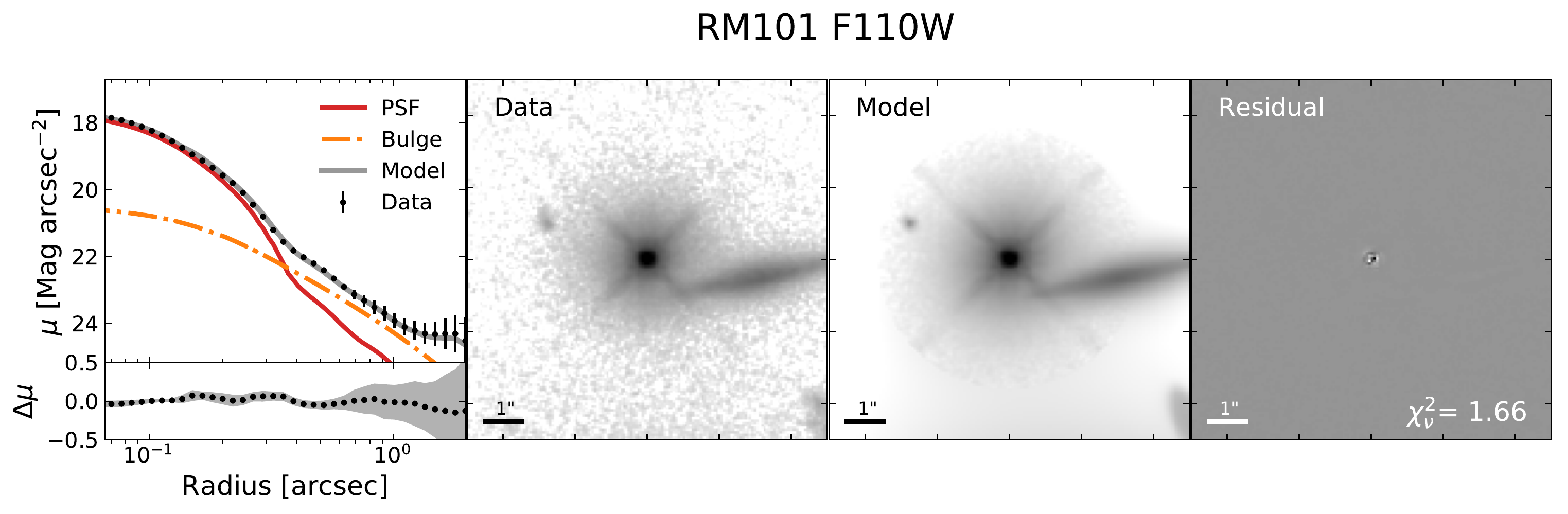}\\
\includegraphics[width=0.9\textwidth]{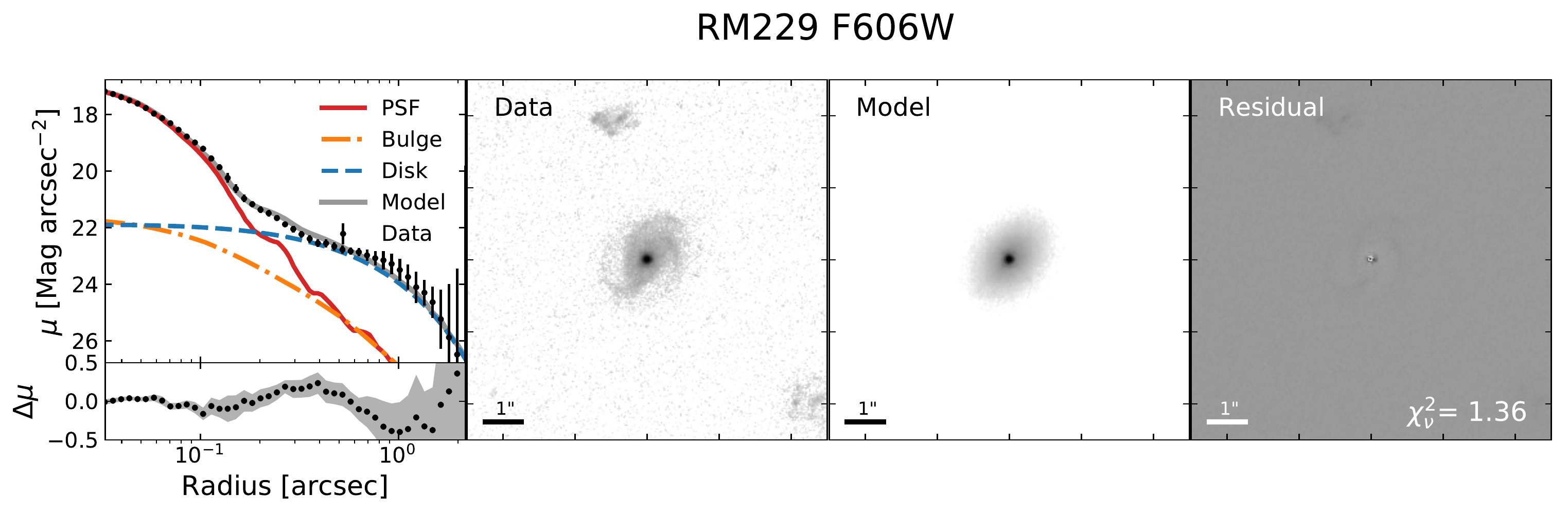}\\
\includegraphics[width=0.9\textwidth]{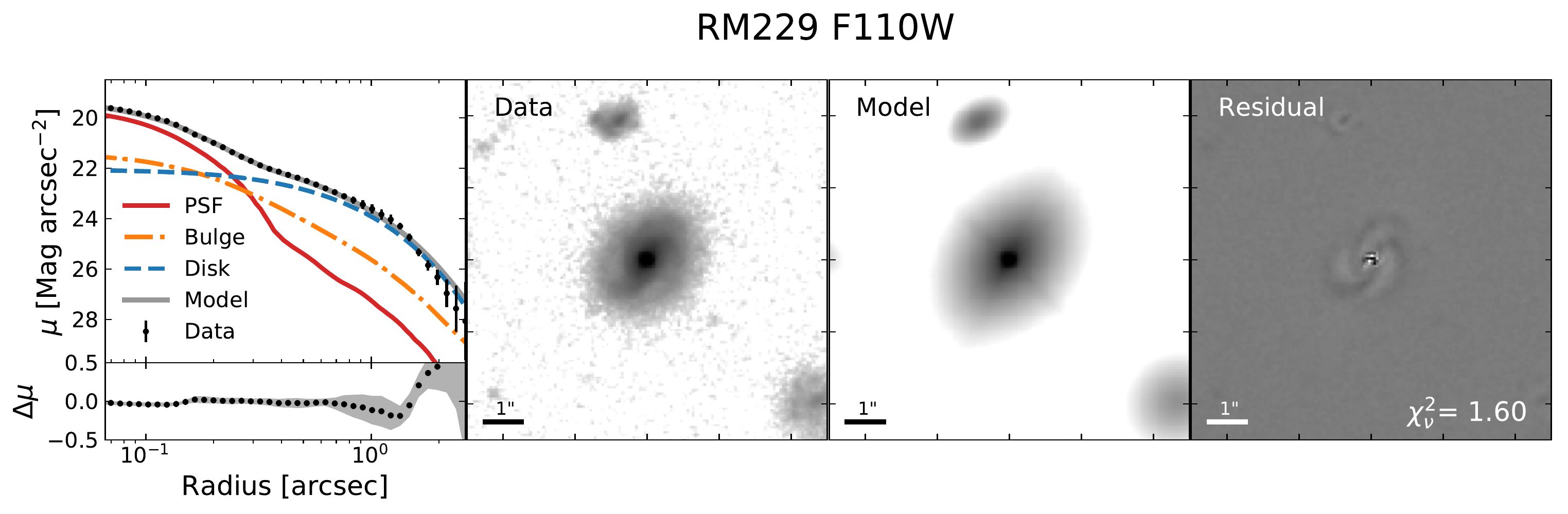}\\
\caption{ 
Surface brightness decomposition of all sources in F606W and F110W bands. The left panel is the surface brightness profile of the data (black dots), the model (grey solid line) and each modeled component (red solid lines for PSFs, orange dotted-dash lines for hosts/bulges (n=4), blue dash lines for exponential disks (n=1), and purple dotted lines for truncated rings (RM775)). The radial profiles are directly measured from the GALFIT decomposed models and the HST images with isophote fitting.
The bottom sub-panel (in the leftmost panel) is the residual of the surface brightness profile, with rms along the elliptical path plotted in grey shaded area. The right three images are (from left to right) the HST image, the GALFIT model and the residual. The reduced ${\chi}^{2}$ of the model is labeled in the lower right corner of the residual image. }
\label{fig:radial_RM101_RM229}
\end{figure*}

\begin{figure*}
\setcounter{figure}{0}
\centering
\includegraphics[width=0.9\textwidth]{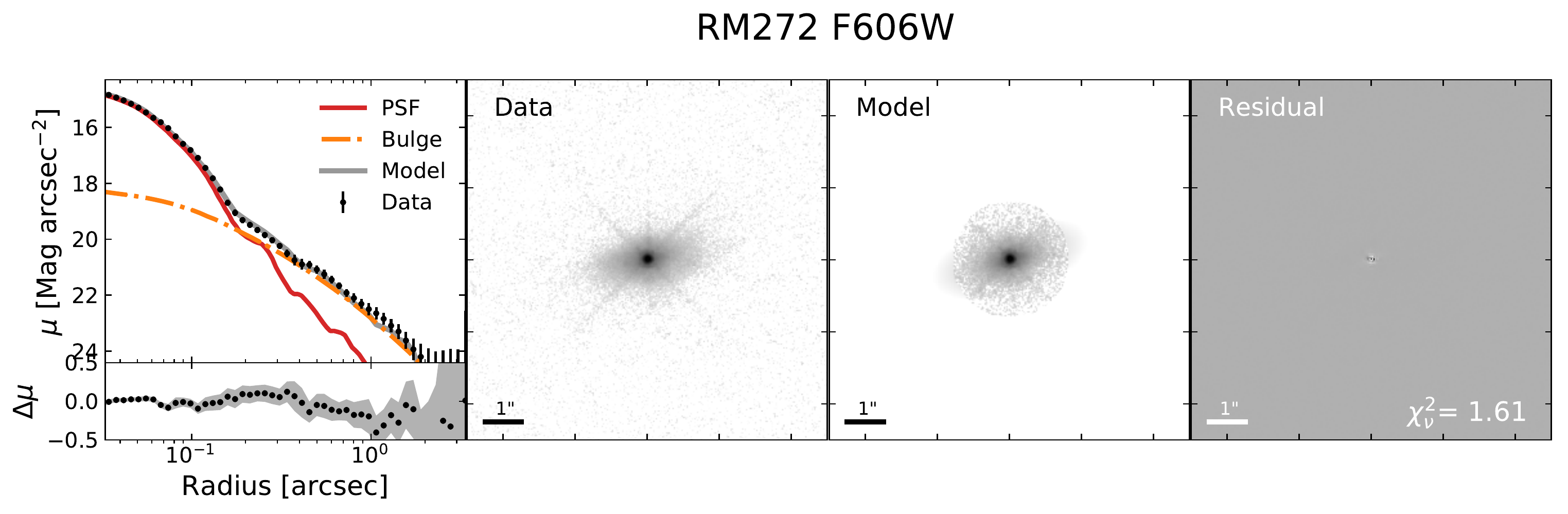}\\
\includegraphics[width=0.9\textwidth]{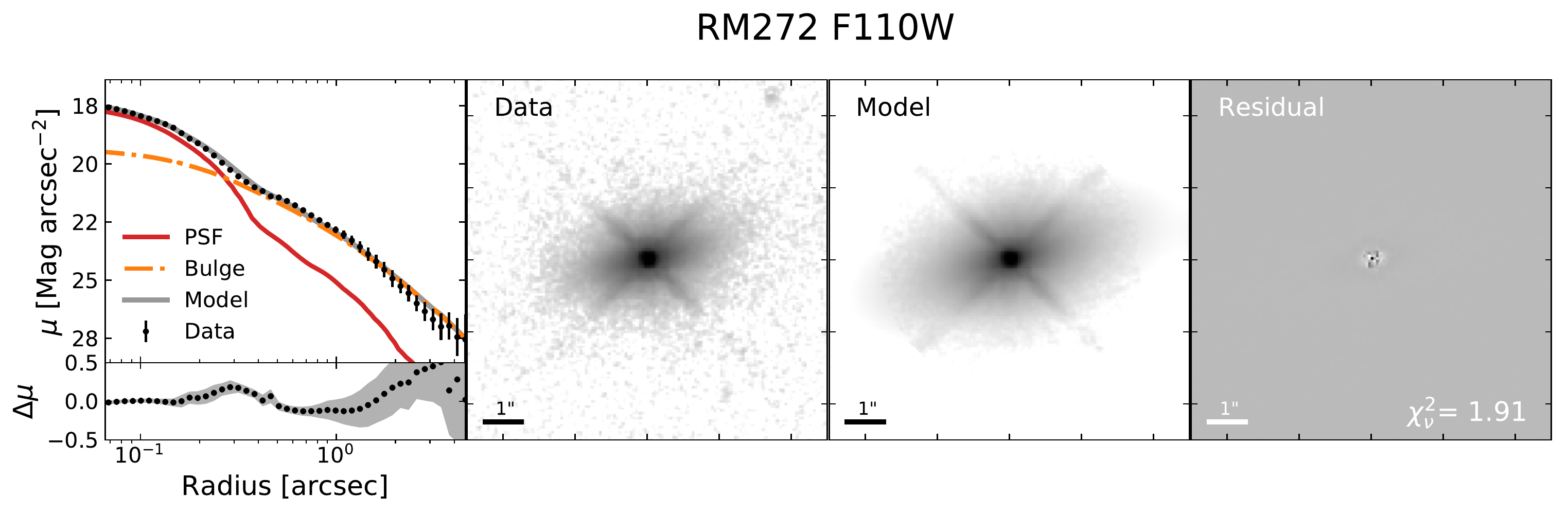}\\
\includegraphics[width=0.9\textwidth]{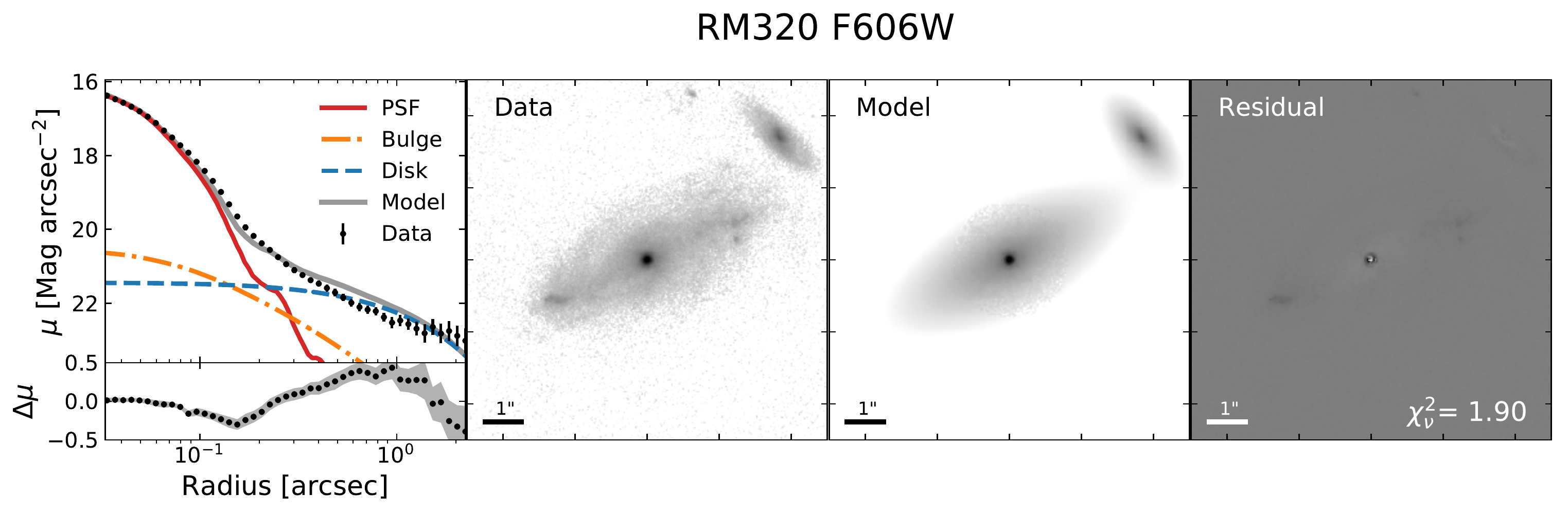}\\
\includegraphics[width=0.9\textwidth]{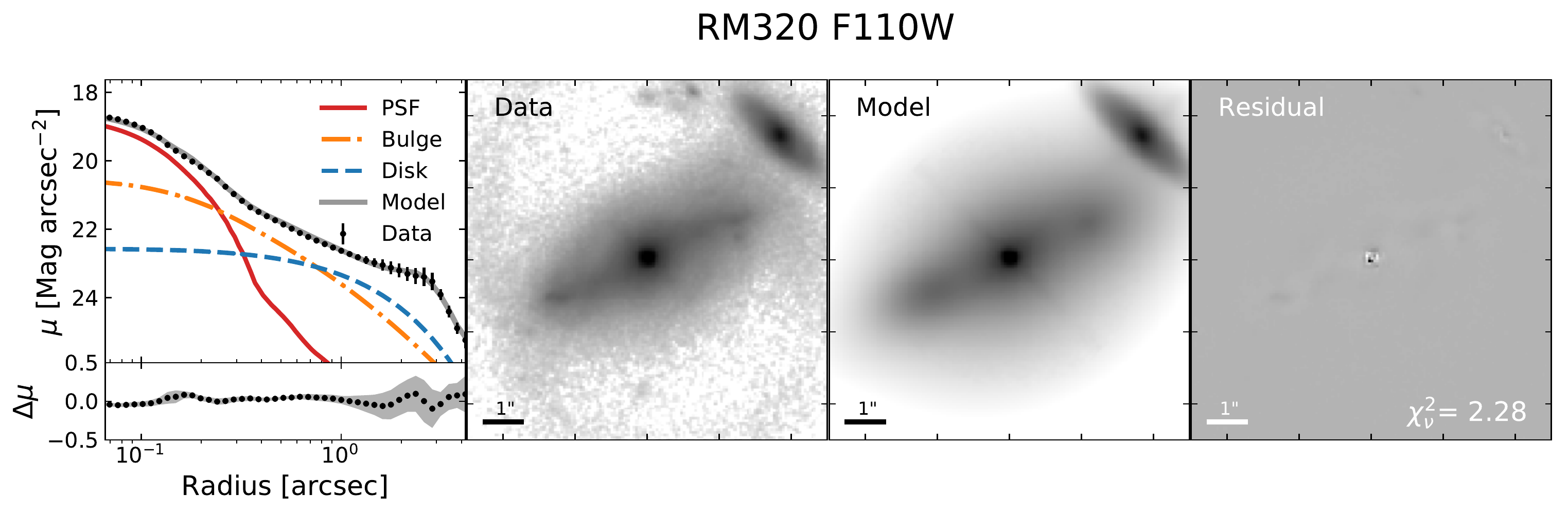}\\
\caption{(continued).}
\label{fig:radial_RM272_RM320}
\end{figure*}

\begin{figure*}
\setcounter{figure}{0}
\centering
\includegraphics[width=0.9\textwidth]{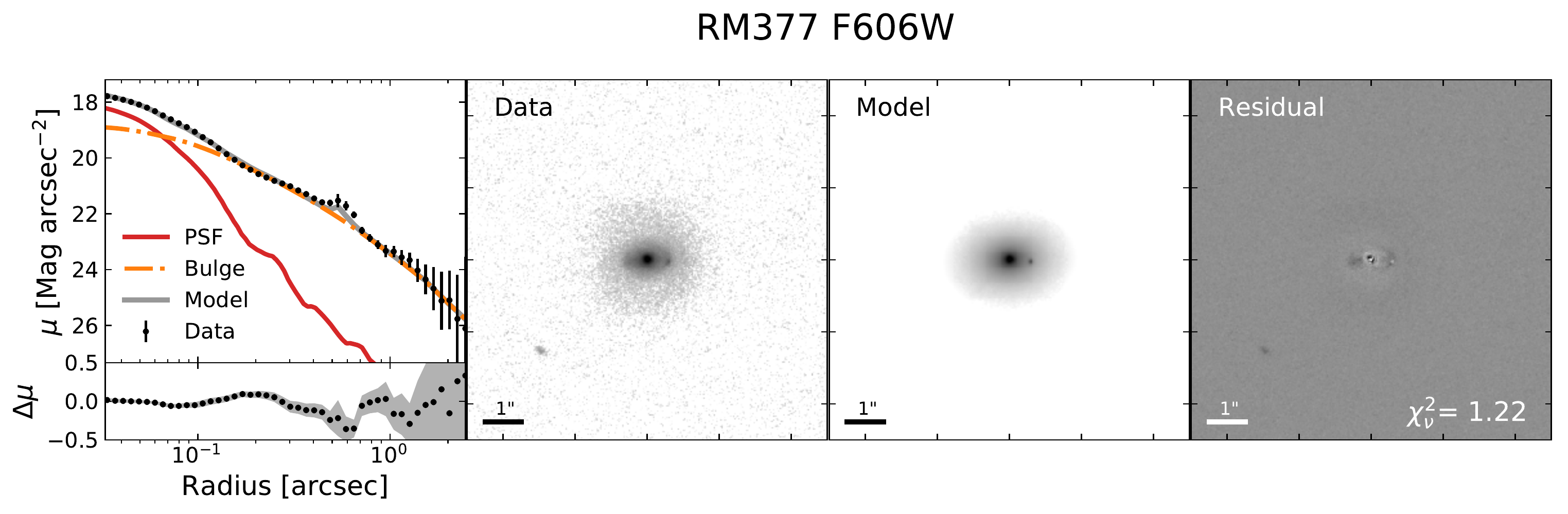}\\
\includegraphics[width=0.9\textwidth]{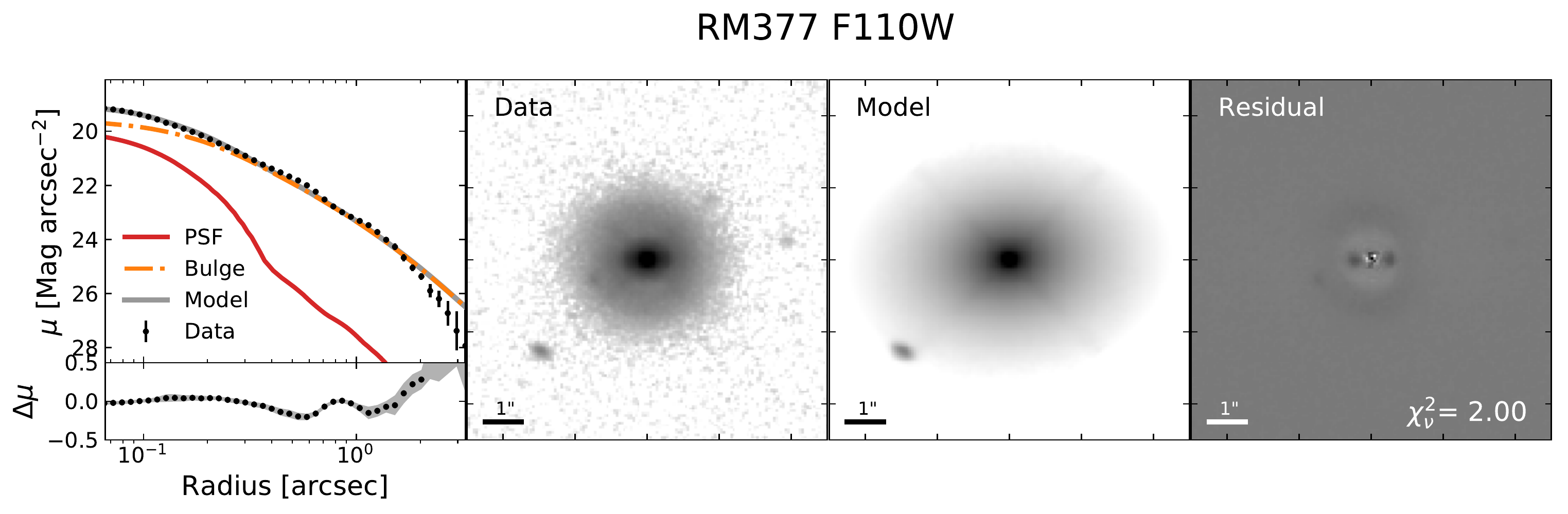}\\
\includegraphics[width=0.9\textwidth]{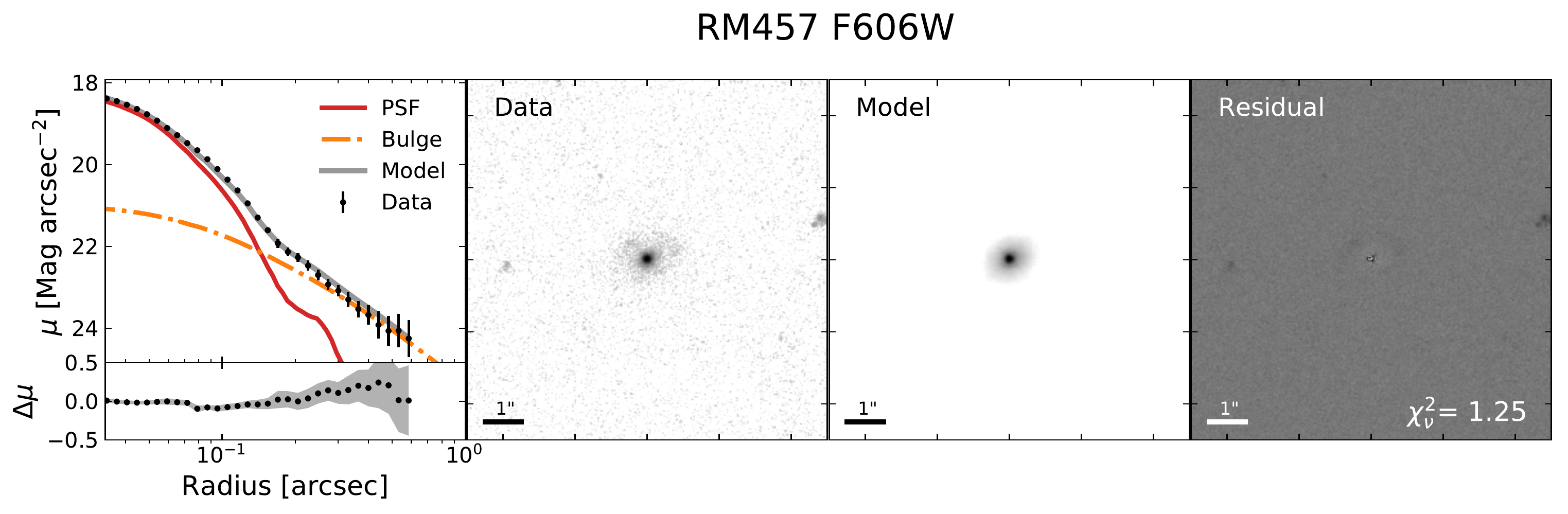}\\
\includegraphics[width=0.9\textwidth]{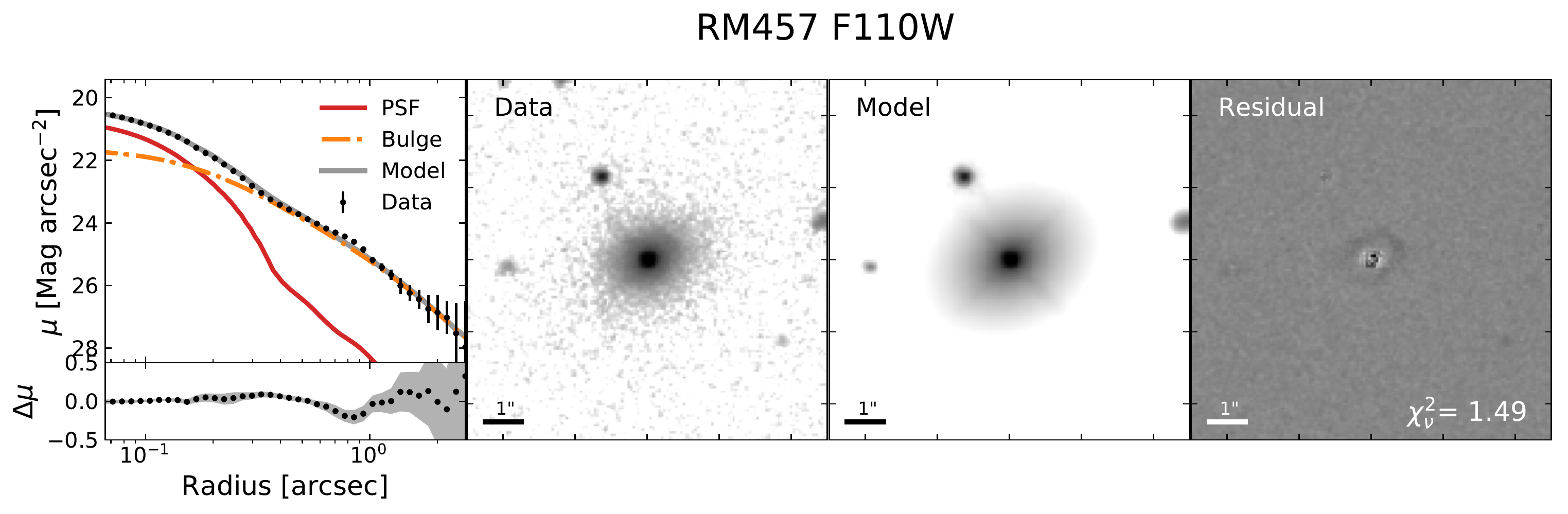}\\
\caption{(continued).}
\label{fig:radial_RM377_RM457}
\end{figure*}

\begin{figure*}
\setcounter{figure}{0}
\centering
\includegraphics[width=0.9\textwidth]{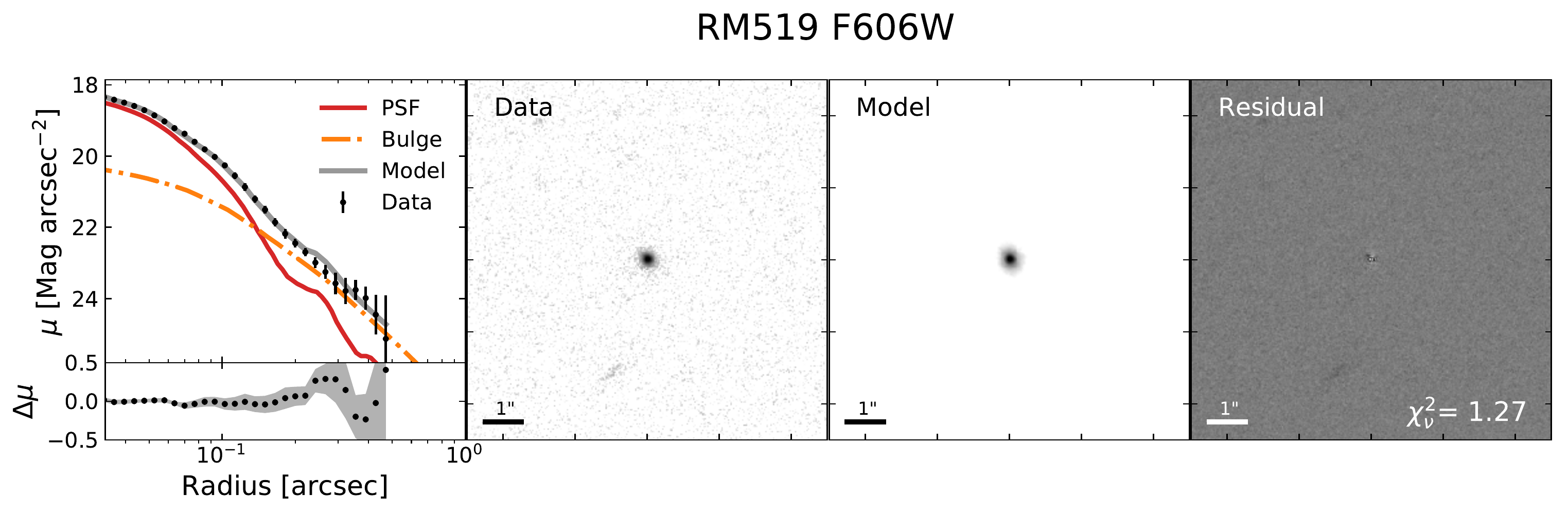}\\
\includegraphics[width=0.9\textwidth]{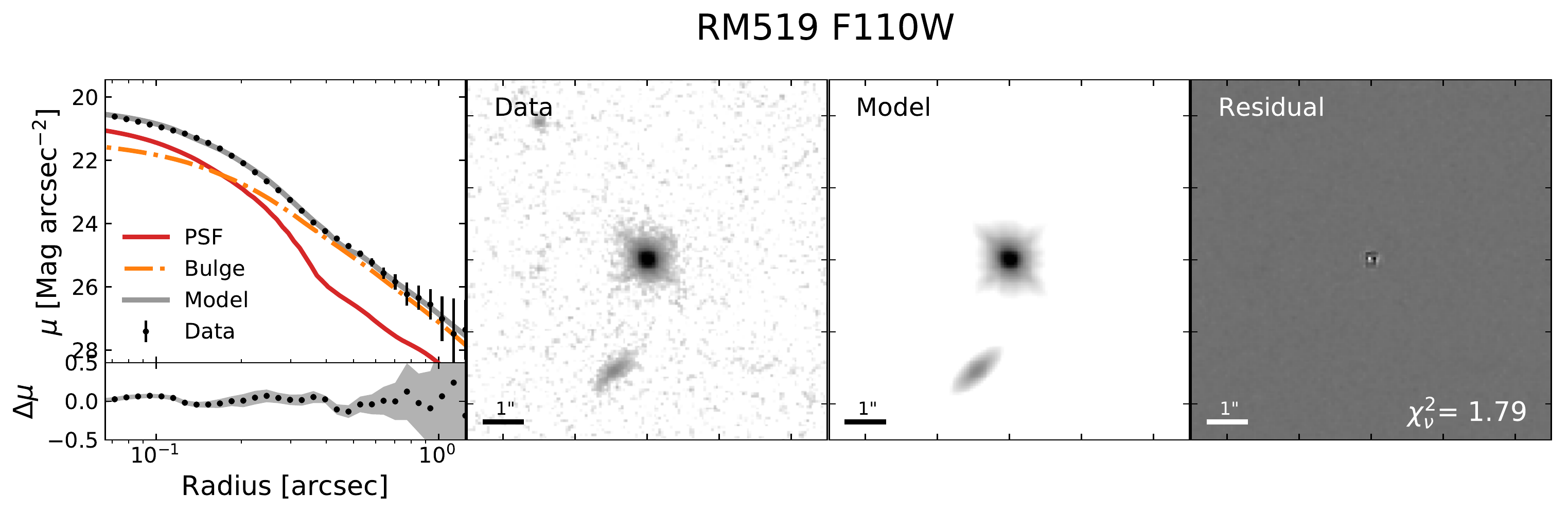}\\
\includegraphics[width=0.9\textwidth]{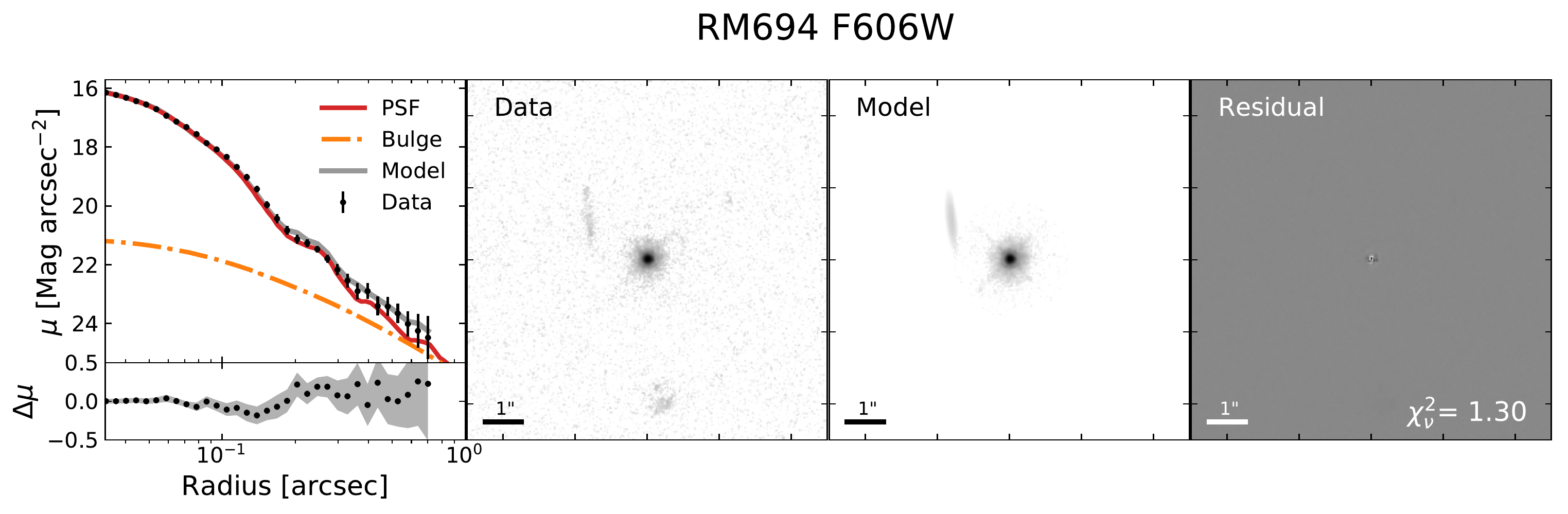}\\
\includegraphics[width=0.9\textwidth]{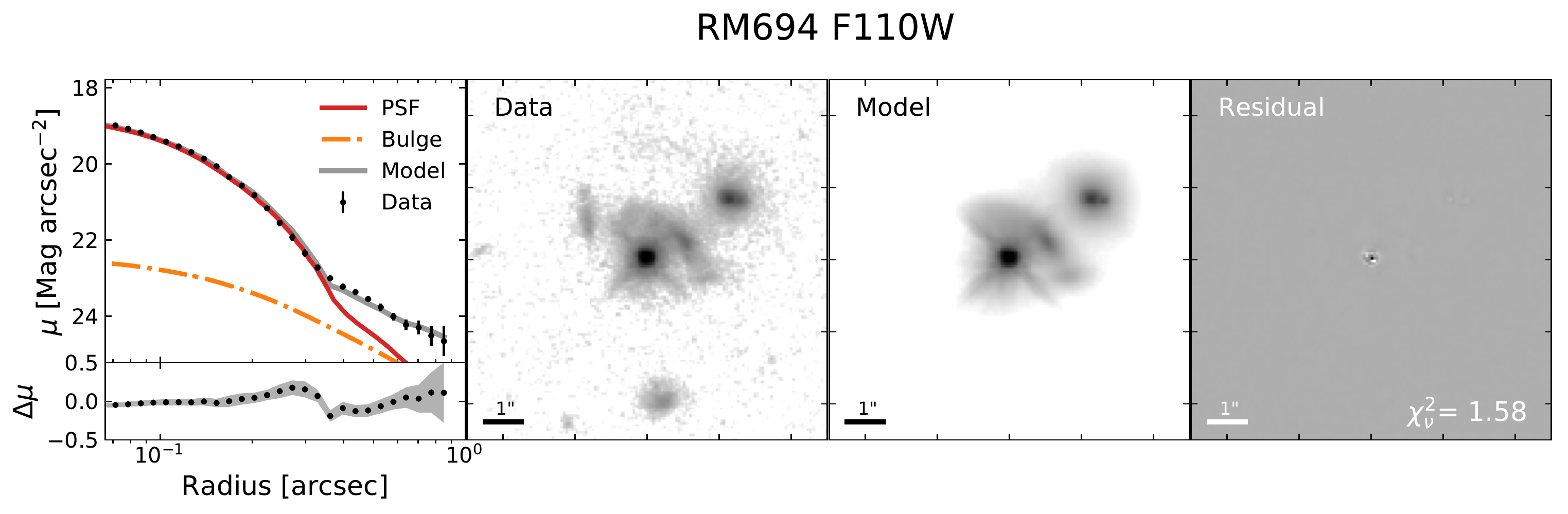}\\
\caption{(continued).}
\label{fig:radial_RM519_RM694}
\end{figure*}

\begin{figure*}
\setcounter{figure}{0}
\centering
\includegraphics[width=0.9\textwidth]{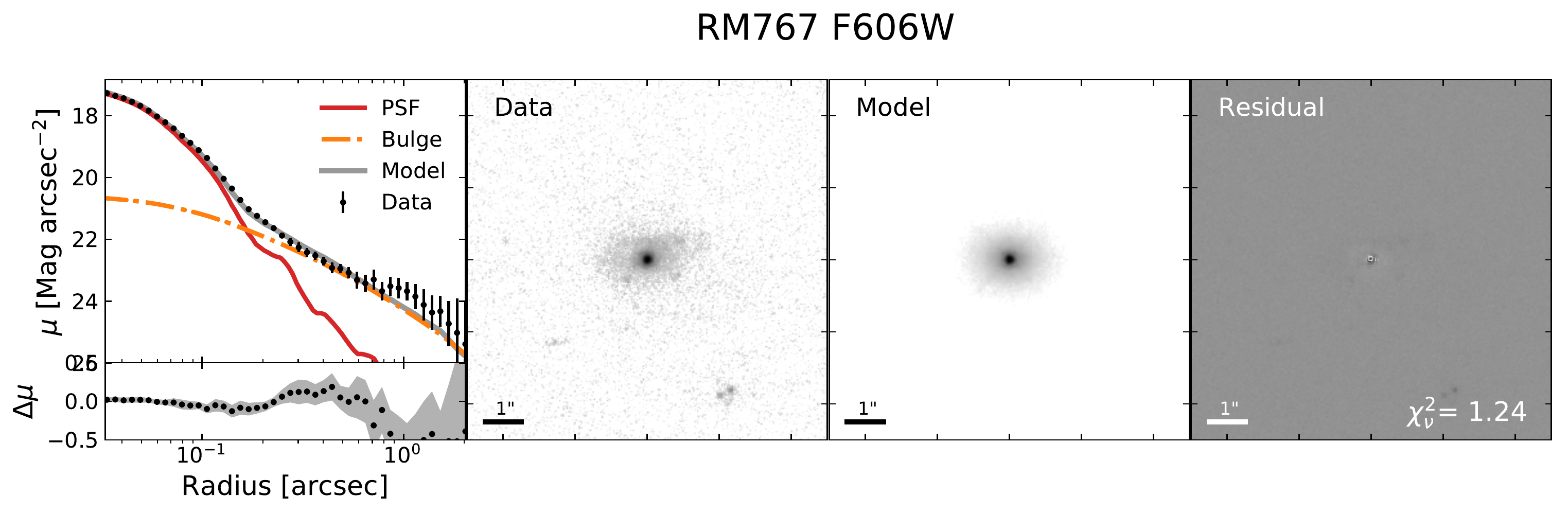}\\
\includegraphics[width=0.9\textwidth]{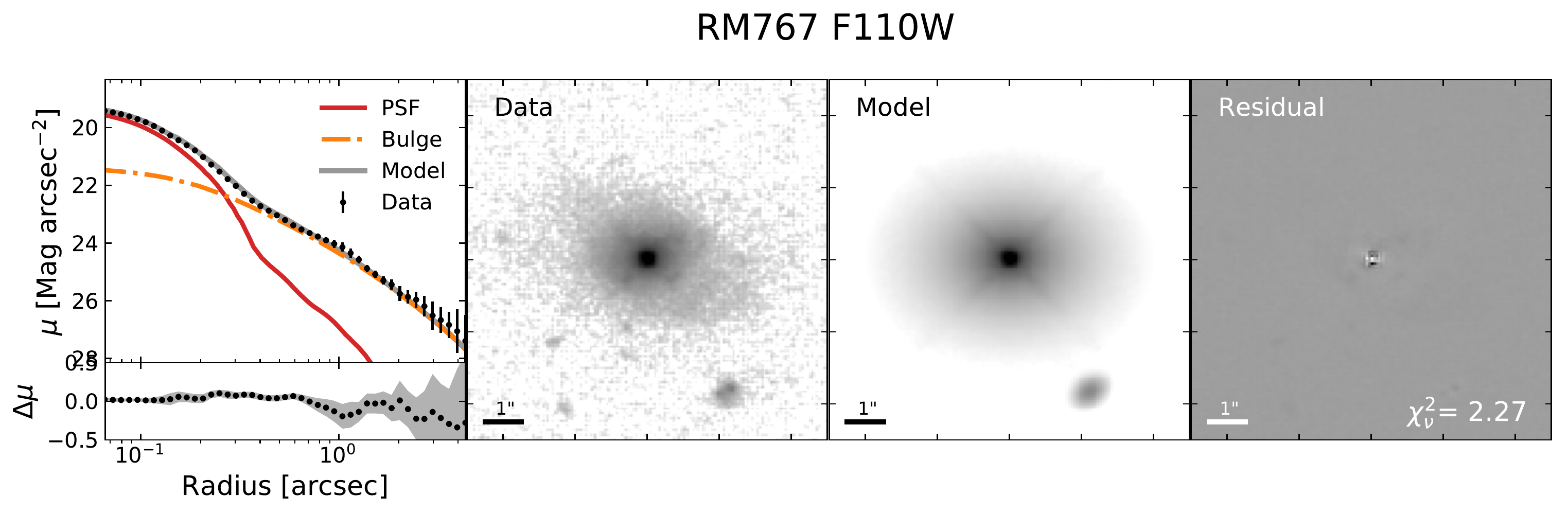}\\
\includegraphics[width=0.9\textwidth]{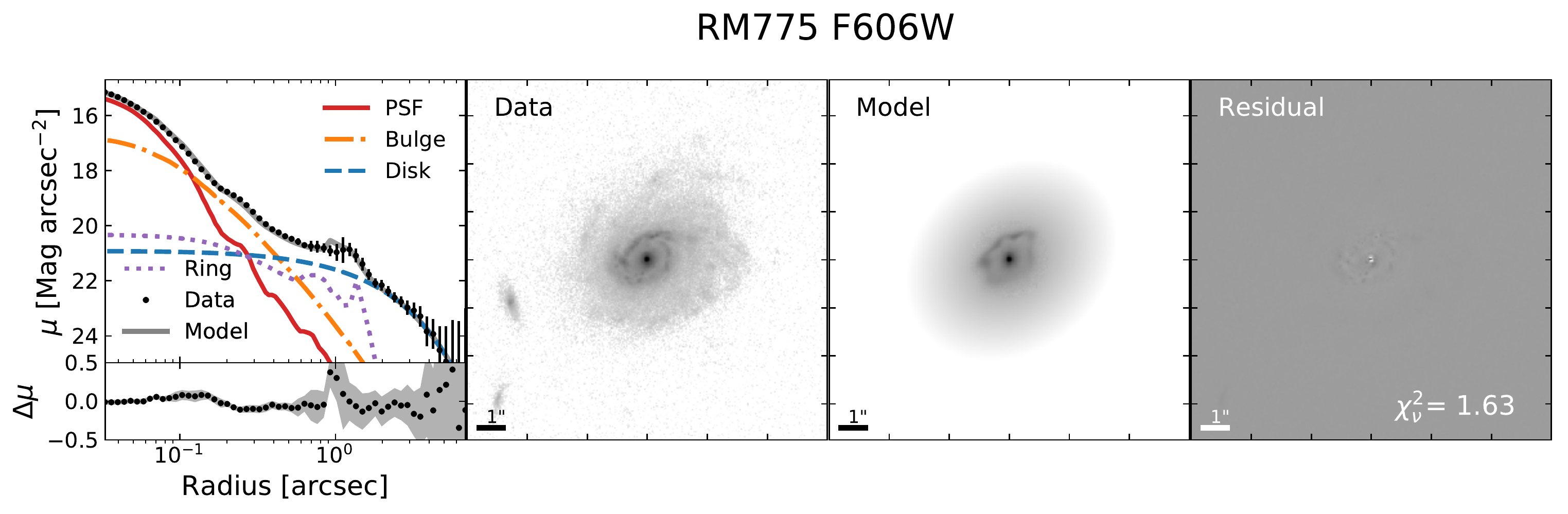}\\
\includegraphics[width=0.9\textwidth]{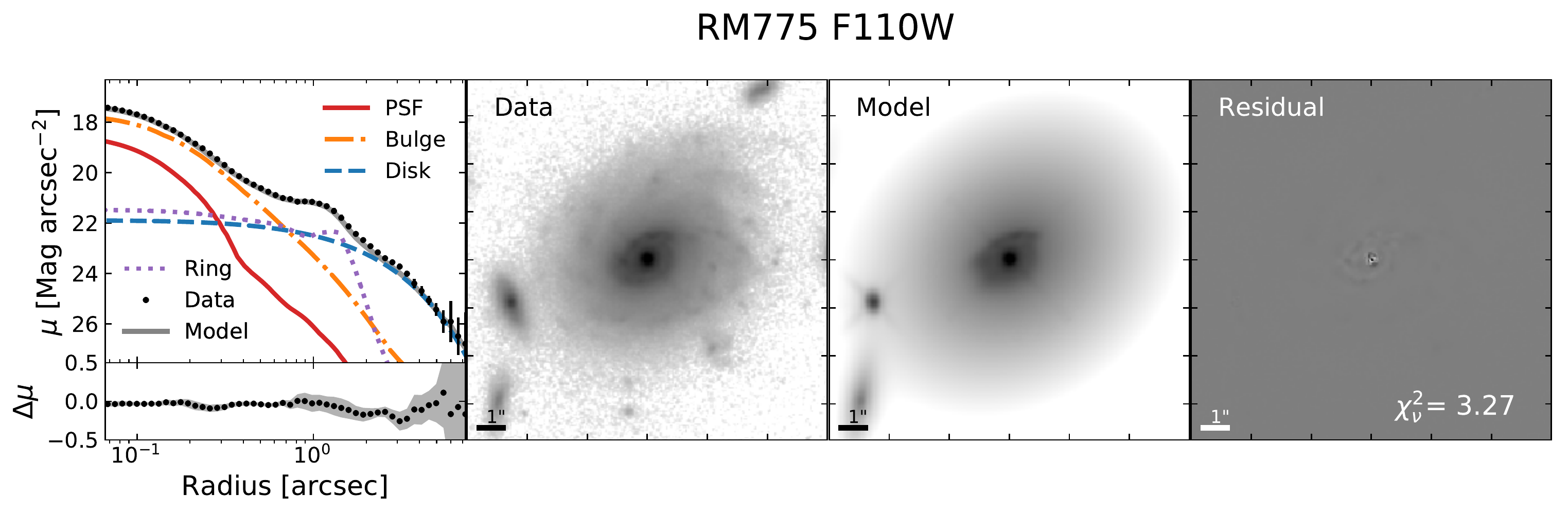}\\
\caption{(continued).}
\label{fig:radial_RM767_RM775}
\end{figure*}

\begin{table*}
\centering
\caption{Galaxy decomposition results}
\setlength\tabcolsep{6pt}
\label{tab:galfit}
\begin{tabular}{c|rrrrrrrrr}
\hline\hline
RMID & Comp. & ${\rm {Mag}_{F606W}}$ & ${\rm {Mag}_{F110W}}$ & r ($''$) & n & q & P. A. & ${\rm {r\chi^{2}}_{F606W}}$ & ${\rm {r\chi^{2}}_{F110W}}$\\
\hline														
101	&	PSF	&	19.41	&	20.57	&		&		&		&		&	1.57	&	1.66	\\
	&	Bulge	&	21.03	&	21.11	&	0.71	&	4	&	0.88	&	-36.1	&		&		\\
\hline																			
229	&	PSF	&	21.49	&	22.51	&		&		&		&		&	1.36	&	1.60	\\
	&	Bulge	&	24.53	&	23.15	&	0.31	&	4	&	0.21	&	-26.0	&		&		\\
	&	Disk	&	21.67	&	21.62	&	0.78	&	1	&	0.69	&	-40.7	&		&		\\
\hline																			
272	&	PSF	&	19.11	&	20.37	&		&		&		&		&	1.61	&	1.91	\\
	&	Bulge	&	20.29	&	20.52	&	0.56	&	4	&	0.42	&	-74.1	&		&		\\
\hline																			
320	&	PSF	&	20.68	&	21.58	&		&		&		&		&	1.90	&	2.28	\\
	&	Bulge	&	21.10	&	20.38	&	1.51	&	4	&	0.82	&	-58.9	&		&		\\
	&	Disk	&	20.08	&	21.18	&	1.77	&	1	&	0.36	&	-63.0	&		&		\\
\hline																			
377	&	PSF	&	22.49	&	22.83	&		&		&		&		&	1.22	&	2.00	\\
	&	Bulge	&	20.46	&	20.40	&	0.67	&	4	&	0.69	&	-85.5	&		&		\\
\hline																			
457	&	PSF	&	22.71	&	23.55	&		&		&		&		&	1.25	&	1.49	\\
	&	Bulge	&	22.38	&	22.21	&	0.80	&	4	&	0.75	&	-59.5	&		&		\\
\hline																			
519	&	PSF	&	22.79	&	23.66	&		&		&		&		&	1.27	&	1.79	\\
	&	Bulge	&	23.45	&	23.46	&	0.13	&	4	&	0.65	&	20.9	&		&		\\
\hline																			
694	&	PSF	&	20.41	&	21.62	&		&		&		&		&	1.30	&	1.58	\\
	&	Bulge	&	23.10	&	23.56	&	0.54	&	4	&	0.54	&	-25.4	&		&		\\
\hline																			
767	&	PSF	&	21.69	&	22.18	&		&		&		&		&	1.24	&	2.27	\\
	&	Bulge	&	21.15	&	21.24	&	1.57	&	4	&	0.73	&	89.6	&		&		\\
\hline																			
775	&	PSF	&	19.69	&	21.38	&		&		&		&		&	1.63	&	3.27	\\
	&	Bulge	&	19.84	&	19.64	&	0.15	&	4	&	0.72	&	-14.6	&		&		\\
	&	Disk	&	18.30	&	19.18	&	2.40	&	1	&	0.80	&	-52.1	&		&		\\
 \hline\hline
 \end{tabular}
 \tablecomments{$r$ is the effective radius of the S\'ersic component, $n$ is the S\'ersic index, $q$ is the ratio between the semi-minor axis and the semi-major axis, and P.A. is the position angle in degrees. The reduced ${\chi}^{2}$ is calculated from the image residual, as reported by GALFIT. Magnitudes are reported in ST magnitude (${\rm mag}_{\rm ST} = - 2.5 \log({F}_{\lambda}) - 21.1$), which is the default output from GALFIT. No extinction corrections are made for these magnitudes. The uncertainties of the GALFIT results are discussed in Section \ref{sec:uncertainities}.}

\end{table*}

\section{Data Analysis}\label{sec:analysis}
\subsection{Surface Brightness Decomposition}
We perform 2-dimensional surface brightness decomposition with GALFIT \citep{galfit}. GALFIT is a package that performs 2D ${\chi}^{2}$-fitting of galaxy images using different functional models, including PSF, S\'ersic profiles and structures such as rings, spiral arms and truncated models.

The PSF model for the F110W images is directly constructed from the calibrated image of the dedicated PSF observation of the white dwarf EGGR-26. However, for unknown reasons\footnote{We have checked other programs that used this specific white dwarf as the PSF observation with similar UVIS filters and dither patterns and did not find this problem. Thus we believe this is not a common failure of our strategy of acquiring a dedicated PSF observation.}, the PSF profiles of EGGR-26 and nearby stars in the dedicated F606W PSF observation are systematically wider than that of the field stars in the target frames. Therefore, instead of using the dedicated F606W PSF observation, we identified isolated field stars in all of the science frames (seven in total), and chose the brightest one to construct the PSF model for image decomposition in all F606W images, which proved to work well. 

Since the IR images are deeper and more host-dominant than the UVIS images, we first perform GALFIT for the F110W images, and use the best-fit parameters as constraints (fixing all structural parameters except for the amplitude) in fitting the UVIS-F606W images. Our fitting procedure starts with fitting the IR image with a PSF component for the quasar, a S\'ersic component for the host galaxy and a flat sky background. Typical bulges have S\'ersic indices ($n$) around 1--4, and typical elliptical galaxies have S\'ersic indices around 3--8 \citep{Gadotti_2009, Huang_etal_2013, Salo_etal_2015, Mendez-Abreu_2017, DallaBonta_etal_2018, Guo_etal_2020}. Extensive simulations by \cite{Kim_etal_2008a} have shown that, in these ranges of S\'ersic indices, fixing $n=4$ recovers the host magnitudes better than allowing $n$ to be a free parameter. Therefore, we fix $n=4$ for the bulge component in all our targets. Image decomposition of nearby ($z<0.3$) AGN has shown that a single S\'ersic component is usually sufficient for decomposing the host from the AGN \citep{Kim_etal_2008b, KimM_etal_2017}. An additional disk component (fixed to $n=1$, i.e., an exponential disk) is added only when there is strong evidence of a disk in the residuals of the image and surface brightness profile. Similarly, \cite{Kim_etal_2008a} have shown that $n=1$ is a reasonable assumption for recovering magnitudes of S\'ersic components with $n<2$. We further discuss the uncertainties originated from fixing the S\'ersic indices in our magnitude measurement in Section \ref{sec:uncertainities}. \cite{Bennert_etal_2010} showed that the bulge contribution tends to be underestimated when fitting more than one S\'ersic components to images with low S/N. Following these earlier studies, we only include the disk component if it significantly improves the fitting (reduced ${\chi}^{2}$ in {GALFIT} improved by more than 0.25). For seven targets, fitting with one point source (quasar light) plus one bulge component is sufficient, and adding a disk component to the fit does not improve the reduced ${\chi}^{2}$ by more than 0.25. We include a disk component for three targets (RM229, 320, 775) in which adding the disk improves the reduced ${\chi}^{2}$ by more than 0.25.

In addition, RM775 shows a prominent asymmetric ring feature at $\sim$1$''$ from its center in the IR image, which cannot be modeled by simple S\'ersic profiles and could bias the host flux measurement if not removed properly. We model this ring component using a $n=1$ disk with a truncated inner edge and Fourier modes enabled by GALFIT.

For the UVIS image, we fix all the shape and structural parameters (S\'ersic index, effective radius, ellipticity and position angle) to the best-fit values from the IR image decomposition, and fit for the fluxes of each component only. While the host galaxy does not necessarily have the exact same shape and profile in the two bands, constraining the host parameters can provide more reasonable results on the bulge measurements in the UVIS band images, especially for sources with dim or compact hosts. We have tested fitting the UVIS images without the constraints from the IR results and found that the magnitudes of the decomposed components are roughly the same as before (typical difference is $\sim0.02$\,mag in PSF magnitude and $\sim0.2$\,mag in bulge and disk magnitudes). However, relaxing these constraints often results in structural parameters (such as the S\'ersic index) reaching the limits of GALFIT. Therefore we report our fiducial UVIS decomposition results with the constrained fits. 

Figure \ref{fig:radial_RM101_RM229} shows the HST images and the best-fit GALFIT results. Table \ref{tab:galfit} summarizes the best-fit parameters from GALFIT. 

\subsection{Flux Uncertainties}\label{sec:uncertainities}

The flux uncertainties output by GALFIT are usually very small ($<0.02\,$mag) as GALFIT treats the difference between data and model as purely statistical, and does not consider deviations from the model due to more complex galaxy structures, non-uniform sky background or PSF mismatches, etc. \citep{galfit}. To estimate the true uncertainties of the GALFIT magnitudes, we measure the total flux directly from the HST images within an ellipse including the entire host galaxy \citep[determined by isophote fitting with {\tt photutils},][]{photutils} and compare with the total GALFIT magnitudes. We adopt the median difference between the isophote fitting magnitude and GALFIT magnitude as our flux uncertainty from GALFIT, which is $\sim0.06\,$mag in F606W and $\sim0.07\,$mag in F110W for the total (host+quasar) magnitude. 

\begin{table*}[]
\centering
\caption{Final photometry, color, luminosity and stellar mass}
\label{tab:results}
\begin{tabular}{cc|cccccccc}
\hline\hline
RMID & Bands & Comp & ${\rm {m}_{B}}$    & ${\rm {m}_{I/R}}$  & Color & $\log L_{\rm B}$          & $\log L_{\rm I/R}$        & $\log M_{*}$ & $\log M_{\rm *,CIGALE}$       \\
 &  & & [mag]    & [mag]  & [mag] & [${\rm {L}_{Sun}}$]          & [${\rm {L}_{Sun}}$]        & [${\rm {M}_{Sun}}$] & [${\rm {M}_{Sun}}$]       \\
\hline
101	&	B,R	&	Host	&	20.89	&	20.12	&	0.77	&	10.96	$\pm$	0.10	&	10.54	$\pm$	0.10	&	10.44	$\pm$	0.35	&	10.30	$\pm$	0.33	\\
229	&	B,R	&	Host	&	21.43	&	20.57	&	0.86	&	10.81	$\pm$	0.10	&	10.39	$\pm$	0.10	&	10.37	$\pm$	0.35	&	10.30	$\pm$	0.34	\\
	&		&	Bulge	&	24.23	&	22.63	&	1.60	&	9.99	$\pm$	0.10	&	9.57	$\pm$	0.10	&	10.24	$\pm$	0.35	&	9.90	$\pm$	0.42	\\
	&		&	Disk	&	21.61	&	20.52	&	1.09	&	10.83	$\pm$	0.10	&	10.41	$\pm$	0.10	&	10.60	$\pm$	0.35	&	10.16	$\pm$	0.34	\\
272	&	B,I	&	Host	&	20.39	&	19.39	&	1.00	&	10.69	$\pm$	0.10	&	10.14	$\pm$	0.10	&	9.80	$\pm$	0.27	&	10.02	$\pm$	0.34	\\
320	&	B,I	&	Host	&	19.87	&	18.75	&	1.12	&	10.96	$\pm$	0.10	&	10.41	$\pm$	0.10	&	10.14	$\pm$	0.27	&	10.25	$\pm$	0.34	\\
	&		&	Bulge	&	21.36	&	19.38	&	1.98	&	10.71	$\pm$	0.10	&	10.15	$\pm$	0.10	&	10.51	$\pm$	0.27	&	10.27	$\pm$	0.39	\\
	&		&	Disk	&	20.06	&	19.84	&	0.23	&	10.52	$\pm$	0.10	&	9.97	$\pm$	0.10	&	9.08	$\pm$	0.27	&	9.60	$\pm$	0.29	\\
377	&	B,I	&	Host	&	20.49	&	19.24	&	1.26	&	11.00	$\pm$	0.10	&	10.45	$\pm$	0.10	&	10.29	$\pm$	0.27	&	10.37	$\pm$	0.35	\\
457	&	B,R	&	Host	&	22.03	&	21.21	&	0.82	&	10.82	$\pm$	0.10	&	10.40	$\pm$	0.10	&	10.34	$\pm$	0.35	&	10.20	$\pm$	0.34	\\
519	&	B,R	&	Host	&	23.15	&	22.42	&	0.73	&	10.24	$\pm$	0.10	&	9.82	$\pm$	0.10	&	9.68	$\pm$	0.35	&	9.56	$\pm$	0.33	\\
694	&	B,R	&	Host	&	22.97	&	22.54	&	0.44	&	10.15	$\pm$	0.10	&	9.73	$\pm$	0.10	&	9.31	$\pm$	0.35	&	9.39	$\pm$	0.31	\\
767	&	B,R	&	Host 	&	20.93	&	20.26	&	0.67	&	11.05	$\pm$	0.10	&	10.63	$\pm$	0.10	&	10.43	$\pm$	0.35	&	10.38	$\pm$	0.33	\\
775	&	B,I	&	Host	&	18.15	&	17.15	&	1.00	&	11.18	$\pm$	0.10	&	10.63	$\pm$	0.10	&	10.29	$\pm$	0.27	&	10.44	$\pm$	0.33	\\
	&		&	Bulge	&	20.19	&	18.64	&	1.55	&	10.58	$\pm$	0.10	&	10.03	$\pm$	0.10	&	10.08	$\pm$	0.27	&	10.03	$\pm$	0.36	\\
	&		&	Disk	&	18.12	&	17.44	&	0.69	&	11.06	$\pm$	0.10	&	10.51	$\pm$	0.10	&	9.94	$\pm$	0.27	&	10.24	$\pm$	0.32	\\
\hline\hline
\end{tabular}
\tablecomments{Magnitudes are reported in AB magnitudes, and color refers to either B-I or B-R. The bulge mass sometimes exceeds the total host mass due to limitations in stellar mass estimation with 2-band photometry, as discussed in \S\ref{sec:stellar_mass}. The last column lists the stellar masses estimated with {\tt CIGALE} to compare with our fiducial stellar masses. }
\end{table*}

We also evaluate the uncertainties due to fixing the S\'ersic index in GALFIT. The S\'ersic index is degenerate with other fitting parameters, in particular the flux and effective radius of the S\'ersic component. Therefore, we have chosen to fix the S\'ersic index for the bulge or disk component during our fitting procedure to remove parameter degeneracy and to prevent unphysical fitting results (e.g. $n>10$). For a sanity check, we allow the S\'ersic index to vary in the IR fit. The S\'ersic index converges within $0.9<n<4.4$ for the bulge component (median $n=2.1$) and $0.4<n<2.4$ for the additional disk component (median $n=0.7$) for all but one source, RM320. For RM320, the best-fit S\'ersic index converges to the GALFIT upper bound of $n=20$, which is due to GALFIT attempting to compensate PSF mismatch with a compact host bulge. Comparing the two cases with and without fixing the S\'ersic indices, the central point source fluxes are typically consistent within $\sim$0.03$\,$mag, the bulge and disk fluxes are consistent within $\sim$0.2$\,$mag, which is consistent with the \cite{Kim_etal_2008a} simulations. The effective radii of the bulge and disk components are on average consistent within 15$\%$. 

Combining the flux measurement uncertainties from fitting residuals in images and parameter constraints in the fitting procedure (i.e., fixing the S\'ersic index) in quadrature, we adopt final flux uncertainties of $0.1\,$mag for the quasar component, $0.25\,$mag for the bulge, the disk and the entire host. When disks are present, we still adopt $\sim 0.25\,$mag as the uncertainty for the entire host galaxy, since GALFIT is capable of recovering the total host flux even when the decomposition of the bulge and disk component is ambiguous. These adopted magnitude uncertainties are consistent with the typical uncertainties adopted in previous work based on HST imaging decomposition of quasar hosts \citep[e.g.,][]{Kim_etal_2008b,Jahnke_etal_2009,Bennert_etal_2010,Park_etal_2015,KimM_etal_2017,Bentz_etal_2018} and simulations of similar sensitivity and host/AGN contrast \citep{Kim_etal_2008a}.

\begin{figure}
\centering
\includegraphics[width=0.45\textwidth]{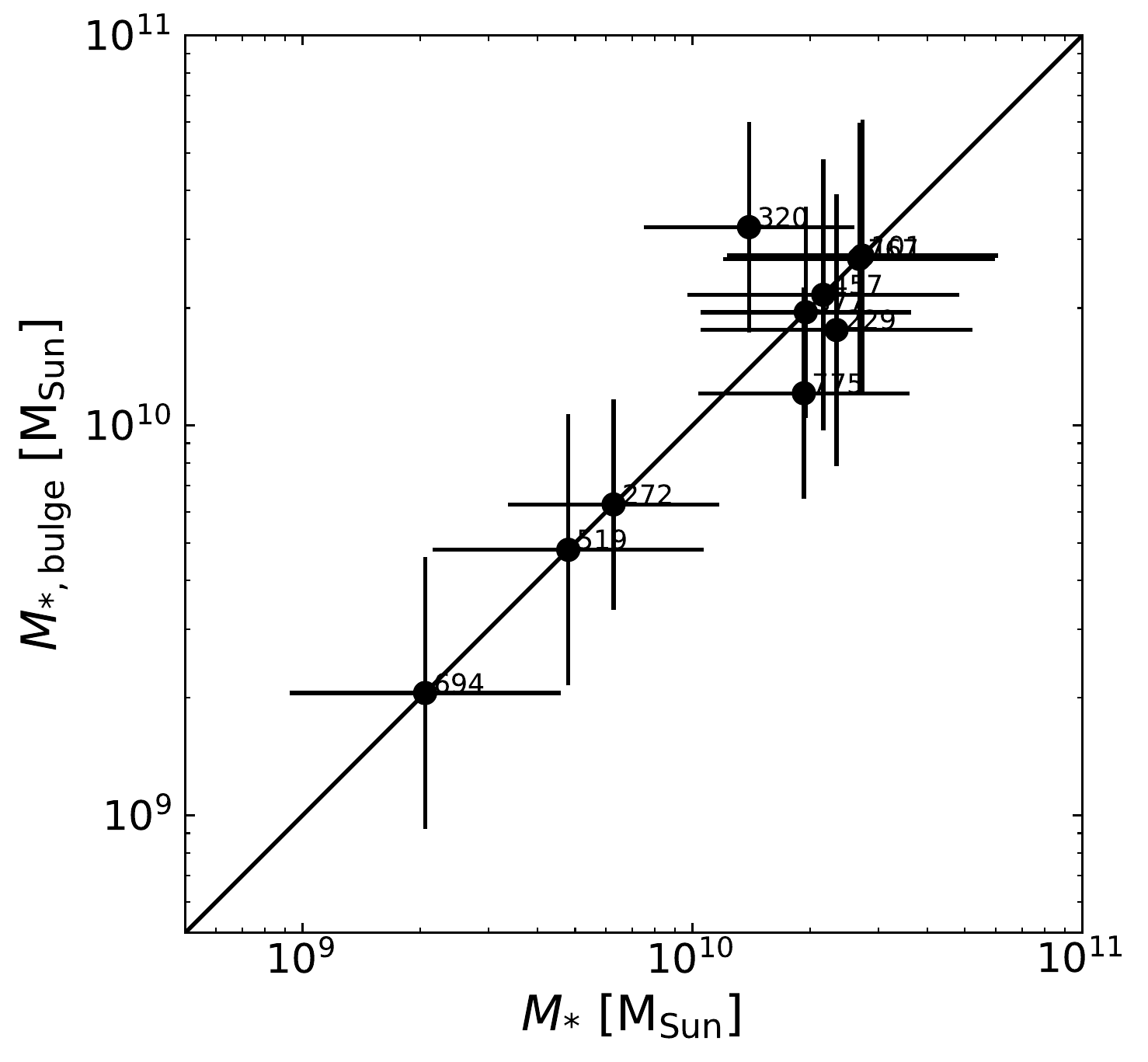}
\caption{Comparison of the stellar masses of the host galaxy (total host mass) and the bulge component. The bulge mass is larger than the total host mass for RM320, which is further discussed in Section \ref{sec:stellar_mass}.}
\label{fig:Mgal_Mbulge}
\end{figure}

\begin{figure*}
\centering
\includegraphics[height=8.1cm]{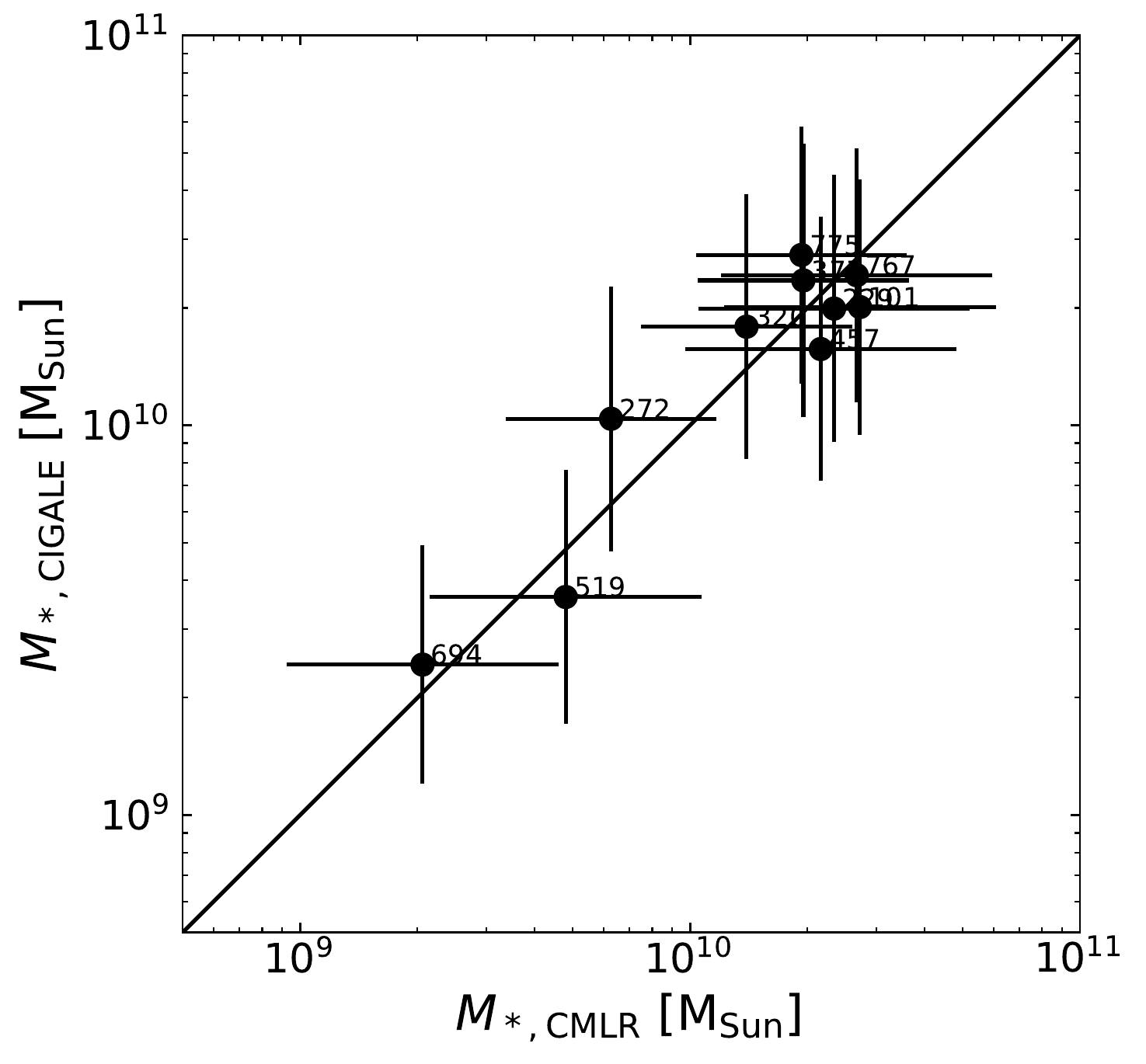}
\includegraphics[height=8cm]{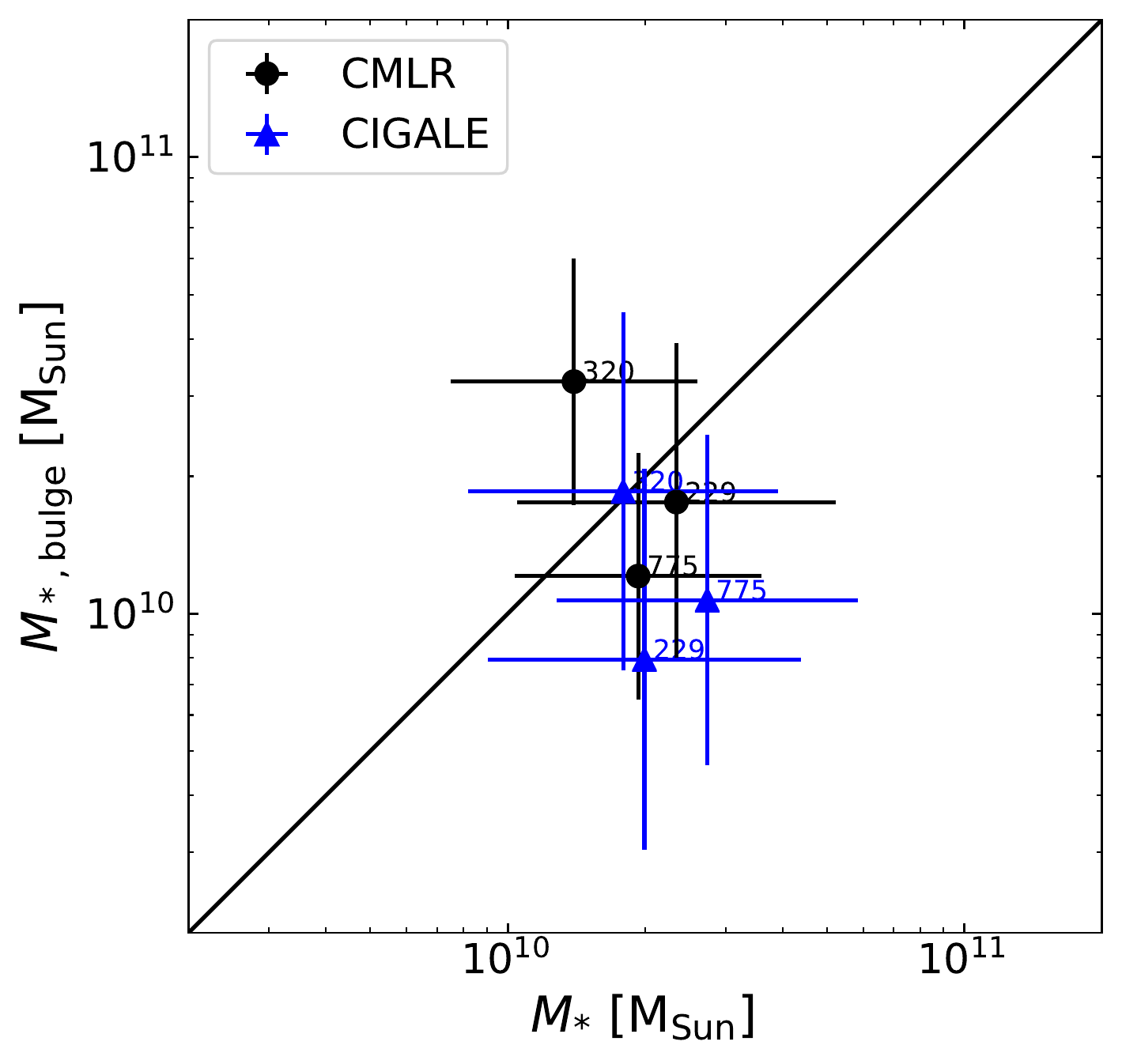}
\caption{Left: Comparison of host stellar masses derived from CMLRs and {\tt CIGALE}. Right: Comparison of the total host stellar mass and bulge mass derived from CMLRs and {\tt CIGALE}.}
\label{fig:cigale}
\end{figure*}

\subsection{Final Photometry}\label{sec:photometry}

To derive the final photometry for our host measurements, we first correct the GALFIT decomposed magnitudes in Table \ref{tab:galfit} for Galactic extinction using the recalibrated \cite{SFD} dust map and reddening in the F606W and F110W bandpasses provided by \cite{Schlafly_etal_2011}. 

To obtain rest-frame photometry, we apply k-corrections and color transformations between the HST filters and Johnson-Cousins filters. We use {\tt CIGALE} \citep{CIGALE} to fit the HST photometry of the hosts with simple population synthesis models, and use the best-fitted spectrum to obtain k-corrections and color corrections in each filter. The {\tt CIGALE} modeling is performed on the bulge and the total host separately if the host is decomposed into a bulge and a disk. We then convert the F606W magnitudes to $B$-band magnitudes for all ten targets. For F110W magnitudes, we convert them to $I$-band magnitudes for sources at $z<0.4$ and to $R$-band magnitudes for sources at $z>0.4$. 
We visually compare the best-fit {\tt CIGALE} model spectra with the decomposed host-only spectra from the SDSS-RM for eight of our targets, as provided by spectral decomposition in \citet{Shen_etal_2015b}, to ensure the {\tt CIGALE} model spectra are reasonable. The SDSS-RM spectra and the {\tt CIGALE} model spectra are generally consistent with each other for compact sources (RM101, RM457, RM519, and RM694), but the {\tt CIGALE} model spectra tend to have more blue flux for more extended sources (RM229, RM320, RM377, and RM775). The SDSS-RM spectra are only from the 2$''$-diameter nucleus region, and do not cover the full wavelength range of the F110W band, so they are not suitable for computing color corrections for the host galaxy. 
The final Galactic-extinction-corrected, k-corrected and band-converted magnitudes for the hosts and bulges are tabulated in Table \ref{tab:results}, which are used for stellar mass estimation in \S\ref{sec:stellar_mass}.

\subsection{Black Hole Masses}
Reverberation mapping measures BH masses by measuring the time delay in variability between the continuum and broad emission lines. The time delay corresponds to the light travel time between the continuum-emitting accretion disk and the Broad Line Region (BLR). Assuming the BLR is virialized, BH masses can be calculated with the time lag ($\tau$) and the width of the broad emission line ($\Delta V$) via the equation: 
\begin{equation}
M_{\rm BH} = f \frac{c {\tau} {\Delta V}^{2}}{G}, 
\end{equation}
where $G$ is the gravitational constant and $f$ is a dimensionless factor that accounts for BLR geometry, kinematics, and inclination. $\Delta V$ can be computed from either the FWHM or the line dispersion $\sigma_{\rm line}$ of the broad line measured from the mean or RMS spectra \citep[e.g.,][]{Wang_etal_2019}. 

Nine of our targets (all except for RM767) have significant \hbeta\ lag detections and RM BH masses from \cite{Grier_etal_2017}. For these nine  sources, we adopt the RM black hole masses from \cite{Grier_etal_2017} computed using a virial coefficient of $f=$1.12 based on FWHM (equivalent to $f=4.47$ when using the line dispersion $\sigma_{\rm line}$ for $\Delta V$). During the first-year of SDSS-RM observations, \cite{Shen_etal_2016a} identified a lag between the continuum and broad \MgII\, line for RM767. However, the lag significance is reduced in the most recent analysis in \cite{Homayouni_etal_2020} using 4-year light curves\footnote{RM767 is not reported in the final significant lag sample in \cite{Homayouni_etal_2020} based on the fiducial lag measurements. We use the \MgII\, lag for RM767 in \cite{Homayouni_etal_2020} based on an alternative approach of lag measurements.}. RM767 is one of the unusual sources that showed more variability in the first-year monitoring, but not the other three years. We use the reported \MgII\ lags for RM767 in both \cite{Shen_etal_2016a} and \cite{Homayouni_etal_2020}, and the broad \MgII\, FWHM from the mean spectrum reported in \cite{Shen_etal_2019b} to estimate the black-hole masses for this work (values reported in Table \ref{tab:sample}). 

\begin{figure*}
\centering
\includegraphics[height=8cm]{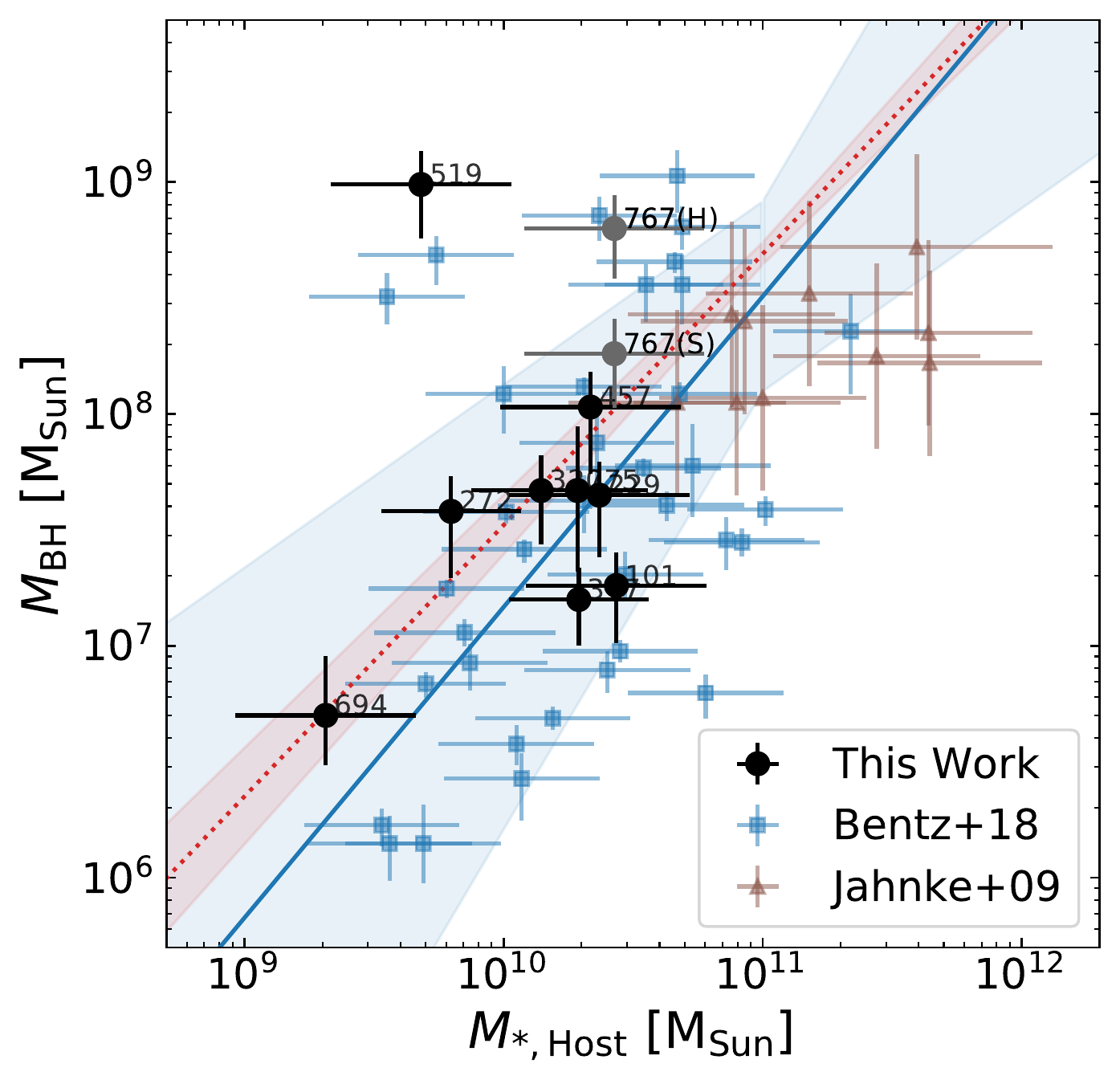}
\includegraphics[height=8cm]{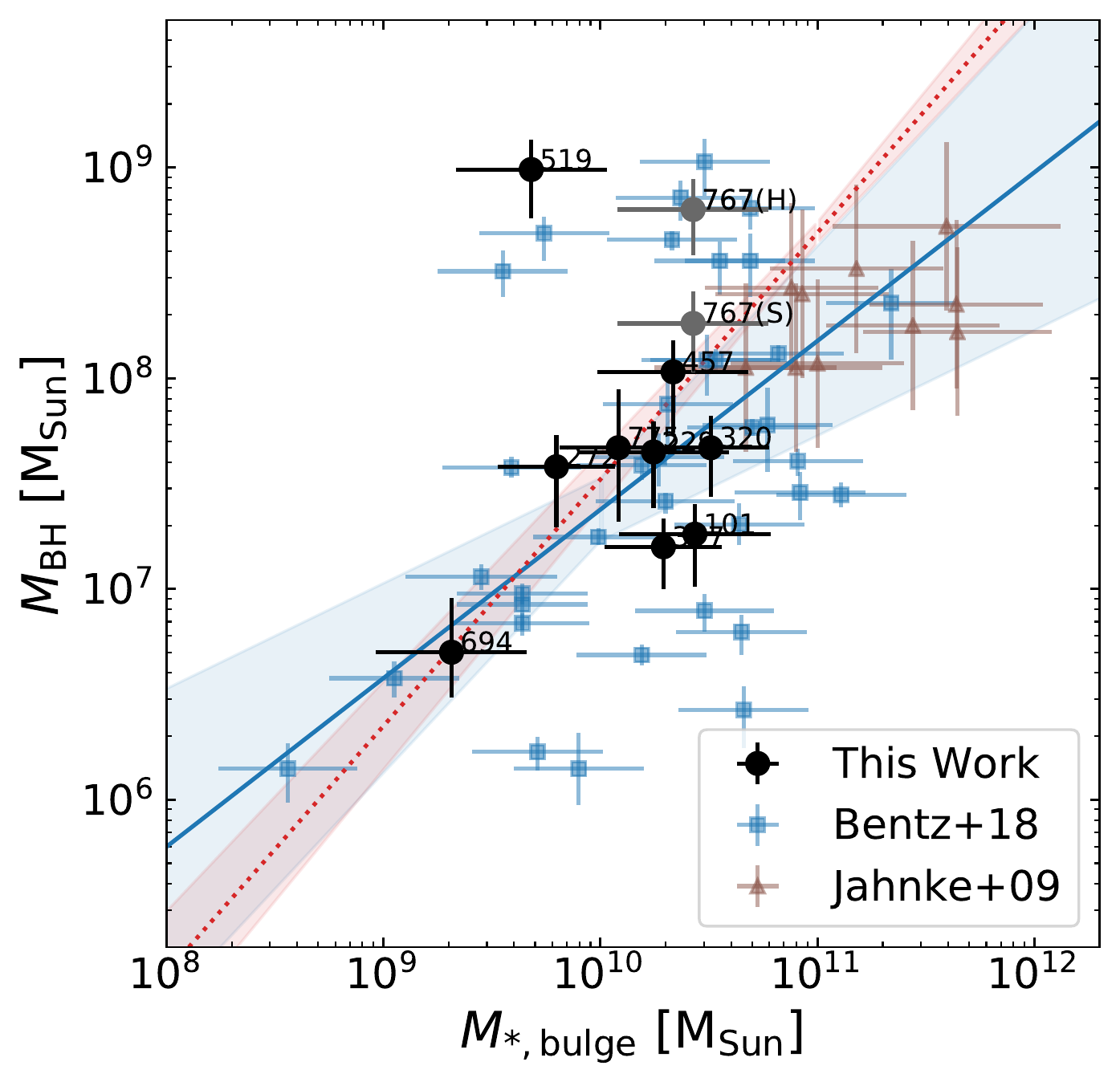}
\caption{Black hole mass as functions of the host-galaxy mass (left) and bulge mass (right). The blue points are the local-RM sample from \cite{Bentz_etal_2018}, and their best-fit relations and $1\sigma$ scatter are shown with blue solid lines and the blue shaded area. The red line denotes the best fit of the black hole mass-bulge mass relation of the local quiescent galaxy sample from \cite{Kormendy_Ho_2013}. For RM767 (grey points), we plot the RM-based BH masses using lags measured from both \citet[][labeled with S]{Shen_etal_2016a} and \citet[][labeled with H]{Homayouni_etal_2020}.} 
\label{fig:Ms_Mgal}
\end{figure*}

\subsection{$M_*$ and $M_{\rm *,bulge}$}\label{sec:stellar_mass}

Following standard practice in the literature \citep[e.g.,][]{Kormendy_Ho_2013,Bentz_etal_2018}, we use the color--${M}_{*}/L$ relations (CMLRs) for dusty galaxy models from \citet[][their Table 6]{Into_Portinari_2013} to derive the bulge and total stellar masses based on 2-band photometry: 
\begin{eqnarray}
    \log_{10}(M_{*}/L_{R})=0.934\times(B-R)-0.832\\
    \log_{10}(M_{*}/L_{I})=0.711\times(B-I)-1.057\ ,
\end{eqnarray}
where colors are rest-frame colors. We apply these CMLRs to the final photometry compiled in Table \ref{tab:results}. We estimate the uncertainties in stellar masses using the propagated uncertainties in photometry. 

\cite{Into_Portinari_2013} constructed the dusty galaxy CMLRs by modeling dust attenuation in a simple spiral galaxy model (i.e., bulge+disk) in various bands following the \cite{Tuffs_etal_2004} prescriptions. CMLRs using optical bands are insensitive to the assumed star formation history and metallicity. However, optical bands are most affected by interstellar dust reddening. To first order, the reddening and extinction effects of dust compensate each other and the CMLRs for dusty galaxies are on average consistent with those for dust-free galaxies \citep{Into_Portinari_2013}. However, the dusty galaxy CMLRs have larger scatter, roughly 0.5\,dex in $\log(M/L)$ at fixed color. Since the colors of our host galaxies are on the bluer end of the \cite{Into_Portinari_2013} galaxy models, we also calculate the host and bulge mass using the dust-free CMLRs in \citet{Into_Portinari_2013}, and the derived stellar masses are consistent within uncertainties.

Figure \ref{fig:Mgal_Mbulge} compares the derived total host mass and bulge mass (if the host is decomposed into a bulge and a disk). The inferred bulge mass is larger than the host mass in RM320, with large uncertainties in both quantities. This appears to be a generic problem for bulge decomposition, and reflects the limitations of using only two-band photometry and empirical CMLRs to estimate stellar masses. For example, in about half of the sample in \citet{Bentz_etal_2018} the reported bulge mass is larger than the total mass. The colors estimated from photometry for our target carry significant uncertainties, which could lead to an apparently larger bulge mass than the host mass. Other systematics from the decomposition procedure, such as PSF mismatch, likely also contributed to this discrepancy.

We also extract the best-fit stellar masses from {\tt CIGALE}. {\tt CIGALE} models galaxy Spectral Energy Distribution (SED) by building composite stellar populations with simple stellar populations (SSP), star formation history and dust attenuation and emission model, using the same IMF and SSP models as in \citet{Into_Portinari_2013}. Figure \ref{fig:cigale} compares the {\tt CIGALE} stellar masses with those from using the CMLRs in \citet{Into_Portinari_2013}. The stellar masses derived from both approaches are generally consistent within 1$\sigma$ uncertainties, but the {\tt CIGALE} model produces bulge masses smaller than host masses. To be consistent with earlier studies \citep[e.g.,][]{Bentz_etal_2018,Vulic_etal_2018,Kim_etal_2019} and facilitate direct comparisons, we adopt the CMLR-based bulge and total host stellar masses as our fiducial values, and report the {\tt CIGALE} stellar masses in Table \ref{tab:results} for reference.

\section{Results}\label{sec:results}
\subsection{The $M_{\rm BH}-M_*$ and $M_{\rm BH}-M_{\rm *,bulge}$  Relations}

Figure \ref{fig:Ms_Mgal} shows the relations between stellar mass and BH mass for total stellar mass (left panel) and bulge stellar mass (right panel). We compare our results with the nearby ($z<0.3$) RM AGN sample in \citet{Bentz_etal_2018}. Their RM-based masses are taken from the AGN Black Hole Mass Database \citep{Bentz_Katz_2015} (originally calculated with $f=4.3$, but rescaled to use $f=4.47$ in Figure \ref{fig:Ms_Mgal} to compare with our BH masses). Their best-fit $M_{\rm BH}-M_*$ and $M_{\rm BH}-M_{\rm *,bulge}$ relations plotted in Figure \ref{fig:Ms_Mgal} are based on stellar masses derived using the \cite{Into_Portinari_2013} CMLRs (using $V-H$ color and $H$ band luminosity). Due to the small sample size, we do not fit a linear relation to the 10 SDSS-RM quasars. Our objects generally fall within the same region occupied by this nearby RM AGN sample. At the high BH-mass end, the two exceptions in our sample (RM519 and RM767, if adopting the \cite{Homayouni_etal_2020} black hole mass) and a small subset of the \citet{Bentz_etal_2018} sample significantly deviate from the best-fit relations. Quiescent galaxies with over-massive black holes are also observed in the local universe \citep[e.g.,][]{Kormendy_Ho_2013,Walsh_etal_2015,Walsh_etal_2017}. Their origins are yet to be understood, but they are suspected to be tidally-stripped or an outlier population in the typical BH-galaxy co-evolution scenario.

\cite{Jahnke_etal_2009} measured host masses of ten type-1 AGN at redshift $\sim1.4$ using two-band HST imaging. Due to limited spatial resolution, they can only distinguish the quasar light from the host light, and were unable to distinguish between the disk and bulge components. Their BH masses, which are derived from the single-epoch method, and host masses (or bulge masses if assuming the bulge is dominant) are in good agreement with the low-$z$ $M_{\rm BH}-M_*$ and $M_{\rm BH}-M_{\rm *,bulge}$ relations (Figure~\ref{fig:Ms_Mgal}). 

As shown in Figure \ref{fig:Ms_Mgal}, our sample is also broadly consistent with the $M_{\rm BH}-M_{\rm *,bulge}$ relation derived from local quiescent galaxies in \cite{Kormendy_Ho_2013}. For a fair comparison, we recalibrate $M_{\rm *,bulge}$ using the \cite{Into_Portinari_2013} CMLRs and the tabulated color $(V-K)$ and $K_s$-band bulge luminosity in their selected sample of ellipticals and classic bulges. The derived $M_{\rm *,bulge}$ are systematically smaller than the tabulated values in \citet{Kormendy_Ho_2013}, but consistent within uncertainties. For simplicity, we use their best fit $M_{\rm BH}-M_{\rm *,bulge}$ (their equation 11) as our local baseline in Section \ref{sec:evo}. Our bulge masses are mostly within $\sim2\sigma$ of the predicted values (except for the outlier RM519 at $\sim3.3\sigma$) from the local $M_{\rm BH}-M_{\rm *,bulge}$ relation in quiescent galaxies.

\begin{figure}
\centering
\includegraphics[width=0.45\textwidth]{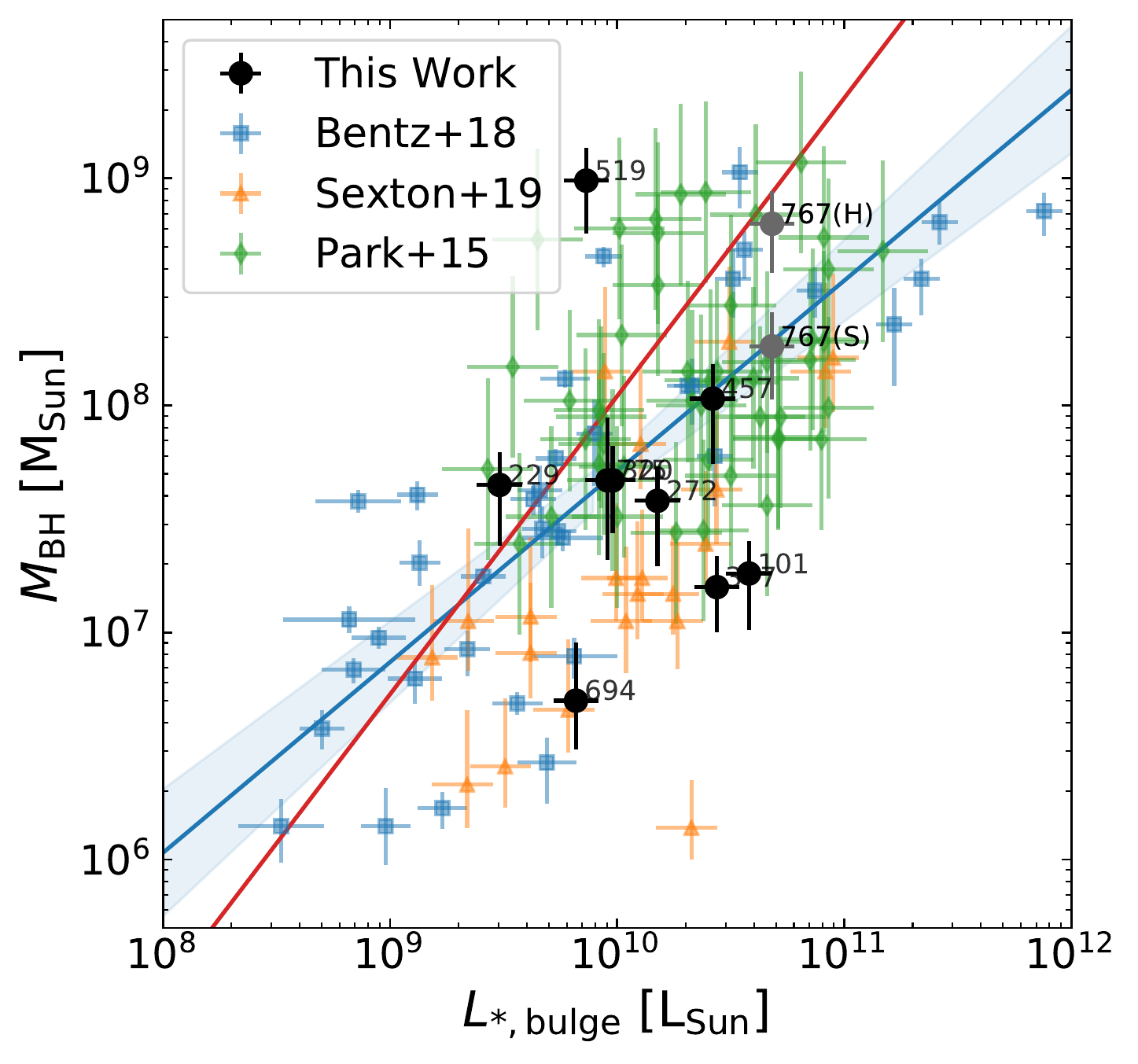} \\
\caption{Black hole mass as a function of bulge luminosity (all in V band, except for \cite{Sexton_etal_2019} in SDSS-r band). Blue line shows the best fit and scatter from the \cite{Bentz_etal_2018} sample and red line shows best fit from the local sample of \cite{Kormendy_Ho_2013}. }
\label{fig:Lbulge}
\end{figure}

\subsection{The $M_{\rm BH}-L_{\rm *,bulge}$ Relation}
The $M_{\rm BH}-L_{\rm *,bulge}$ relation is also a commonly used BH scaling relation. Figure \ref{fig:Lbulge} shows the $M_{\rm BH}-L_{\rm *,bulge}$ relation, along with the local-RM sample \citep{Bentz_etal_2018} and two other samples at intermediate redshifts \citep{Park_etal_2015,Sexton_etal_2019}. The \cite{Park_etal_2015} sample consists of 52 AGN at $z\sim0.36$ and $z\sim0.57$, and the \cite{Sexton_etal_2019} sample consists of 22 AGN in the redshift range of $0.03<z<0.57$. These works both obtained their bulge luminosity through surface brightness decomposition of HST images, and black hole masses are from the single-epoch BH mass estimation. Similar redshift and data quality of the HST images allow us to make direct comparisons among these samples.

\cite{Park_etal_2015} and \cite{Bentz_etal_2018} reported their bulge luminosity in $V$ band. Therefore, we convert our F606W band luminosity to $V$ band luminosity using the best-fit CIGALE SED following the same procedures described in Section \ref{sec:stellar_mass}. \cite{Sexton_etal_2019} reported their bulge magnitudes in SDSS $r$ band, to which we applied a small color correction to $V$ band using galaxy templates of different morphological types provided by \citet{Kinney_etal_1996} and \citet{pysynphot}. This color correction $V-r_{\rm SDSS}$ has values in the range of $0.34-0.55$, with a typical uncertainty of 0.15 from different galaxy templates. As shown in Figure \ref{fig:Lbulge}, all these samples are consistent with the best-fit $M_{\rm BH}-L_{\rm *,bulge}$ relation from \cite{Bentz_etal_2018}, although the scatter is generally large.

For local quiescent galaxies, \cite{Kormendy_Ho_2013} only reported the best-fit $M_{\rm BH}-L_{\rm *,bulge}$ relation in $K_s$ band but not in $V$ band. To compare with our sample and other non-local AGN samples, we use the tabulated $V$-band luminosity and $M_{\rm BH}$ to find a best-fit relation. Our best fit relation has a slightly shallower slope, but is still consistent with the $M_{\rm BH}-L_{\rm *,bulge, K_s}$ relation in \citet{Kormendy_Ho_2013}, with a scatter of 0.22\,dex. We use our best-fit relation as the local baseline in Section \ref{sec:evo}. 

\begin{figure*}
\centering
\includegraphics[width=0.45\textwidth]{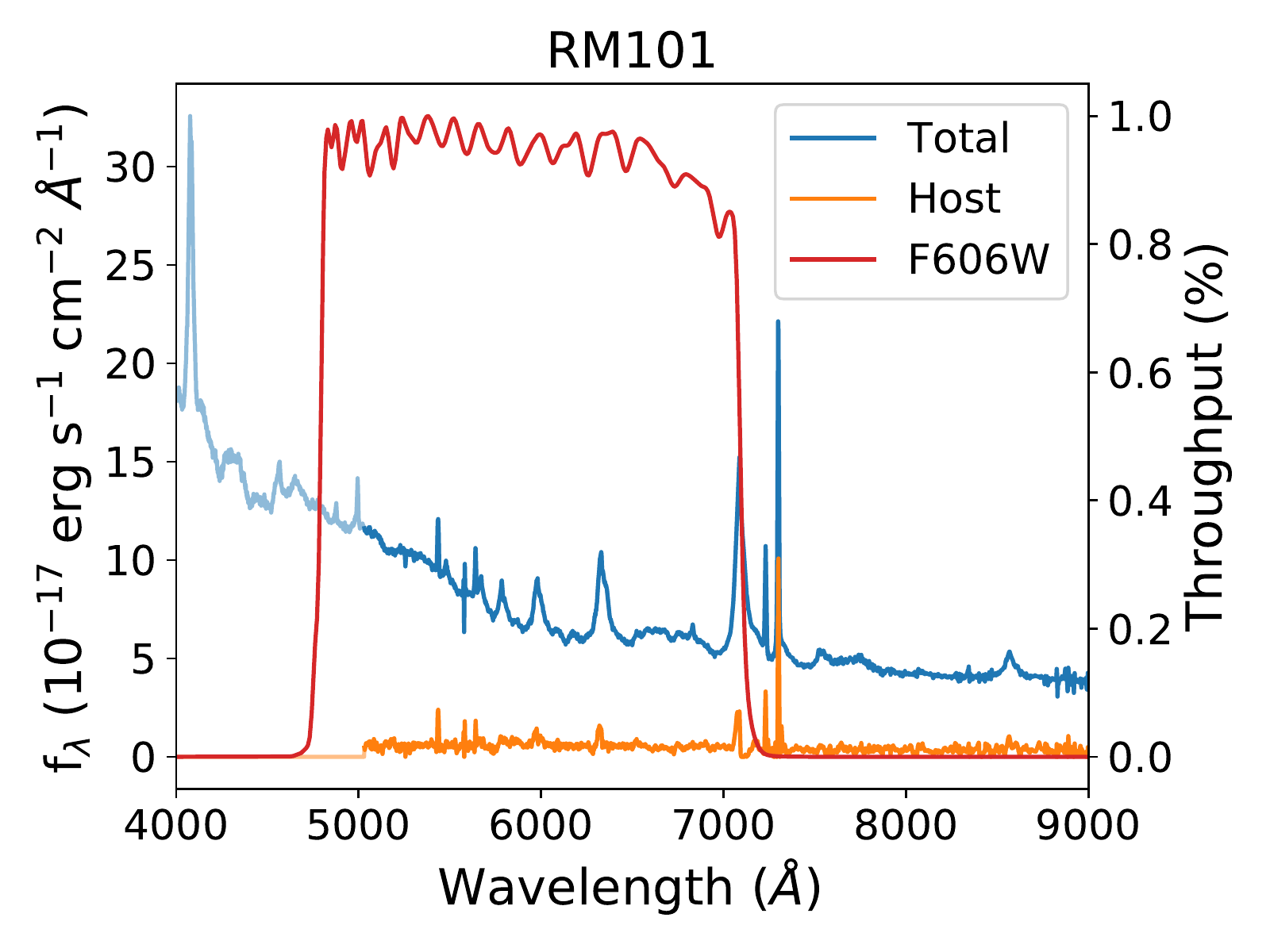}
\includegraphics[width=0.35\textwidth]{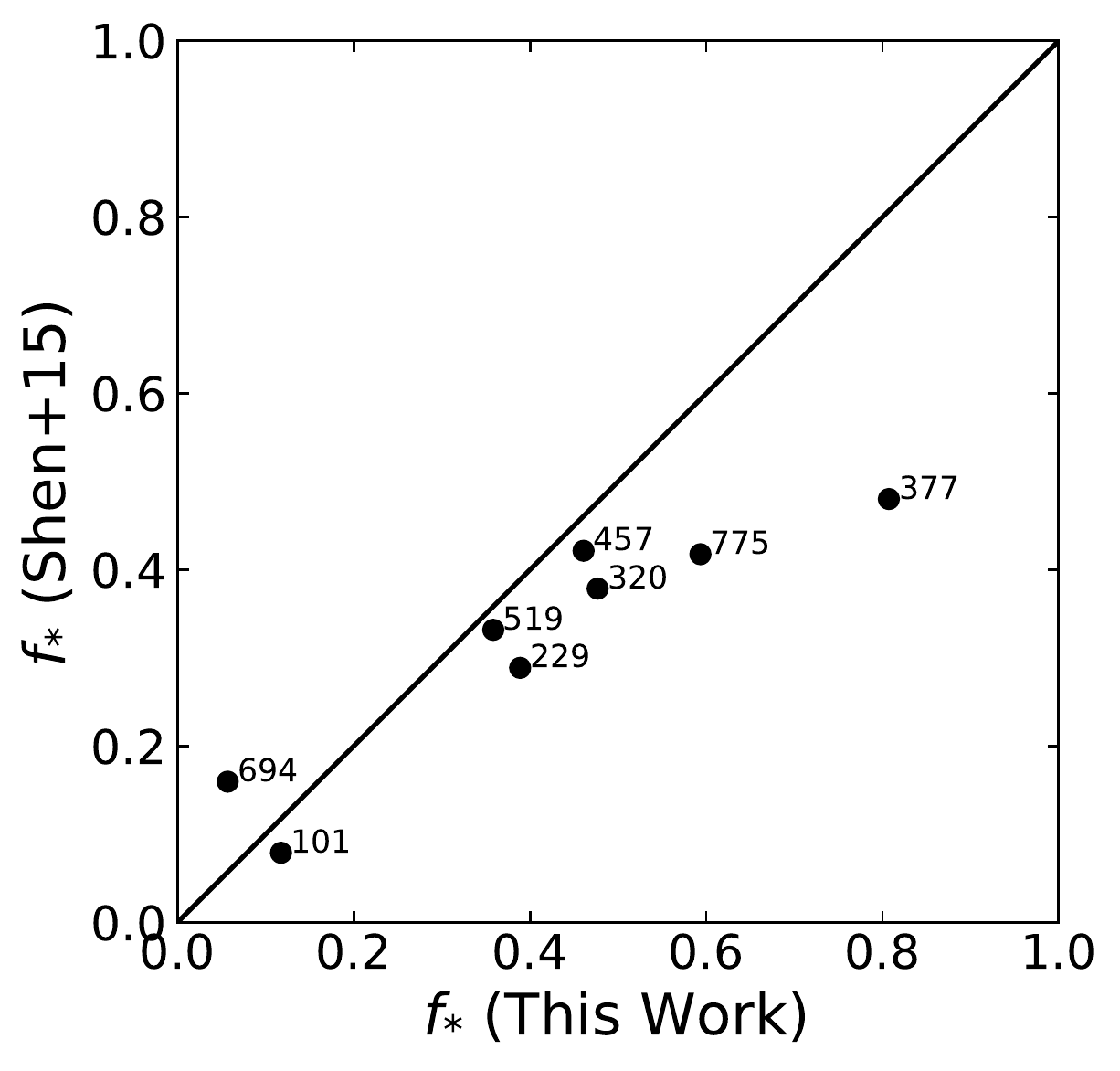}\\
\caption{Comparison of spectral decomposition and image decomposition in the estimation of host fraction in quasars. Left: The HST F606W filter overlaid on the total and decomposed (host only) spectra from \cite{Shen_etal_2015b} in observed wavelength. We only compute the stellar fraction in the spectral range covered by both the total and decomposed spectral, as shown in the thick solid lines. Right: Comparison of the derived stellar fraction from this work and \cite{Shen_etal_2015b} in the F606W bandpass.}
\label{fig:fhost}
\end{figure*}

\begin{figure}
\centering
\includegraphics[width=0.4\textwidth]{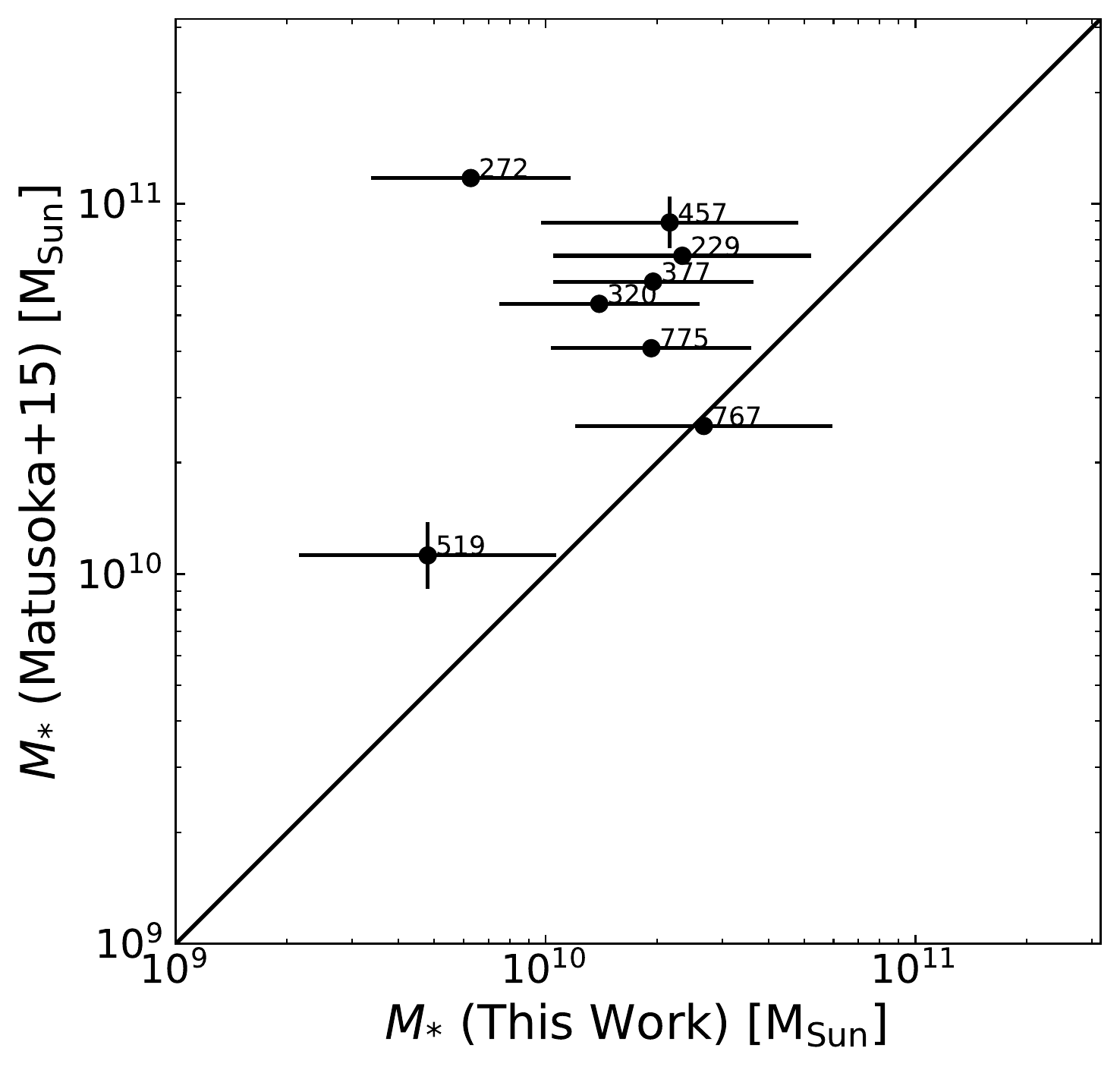} \\
\caption{Comparison of the derived host mass from this work and \cite{Matsuoka_etal_2015}.}
\label{fig:fhost_m15}
\end{figure}

\section{Discussion}\label{sec:disc}
\subsection{Spectral Decomposition versus Image Decomposition}\label{sec:disc_techniques}

For large samples of RM quasars for which HST or other high spatial resolution imaging is unavailable, building a reliable calibration for host properties measured from spectral decomposition is highly desirable.

\cite{Shen_etal_2015b} and \cite{Matsuoka_etal_2015} both measured the host galaxy properties using the high S/N coadded spectra from the first year SDSS-RM monitoring. 
\cite{Shen_etal_2015b} used a Principal Component Analysis (PCA) method to decompose the coadded spectra into the galaxy and quasar spectra to measure stellar properties in quasar hosts, e.g., stellar velocity dispersion, host-free AGN luminosity (at rest frame 5100\,\AA). \cite{Matsuoka_etal_2015} performed spectral decomposition using spectral models of AGN and galaxies. They fit the decomposed galaxy spectra to stellar population models and measured host galaxy properties, including stellar velocity dispersion, stellar mass ($ M_{*}$), and star formation rate. The results from these two works are consistent with each other despite differences in the decomposition technique. 
To evaluate the robustness of spectral decomposition techniques in deriving host properties, we compare the stellar fraction ($f_{*}$, the fractional contribution of the host stellar component to the total flux) from \cite{Shen_etal_2015b} and stellar masses ($M_{*}$) from \cite{Matsuoka_etal_2015} with our HST imaging decomposition results. 

We calculate the stellar fraction from SDSS-RM spectra by computing the expected flux density in the total and decomposed host spectra from \cite{Shen_etal_2015b} in the F606W filter (left panel of Figure \ref{fig:fhost}). When computing the host stellar fraction from our HST imaging decomposition, we only use the decomposed GALFIT models within the 2\arcsec\ diameter spectral aperture. The host fractions from both methods correlate with each other, but the host fraction from spectral decomposition is systematically smaller than that estimated from imaging decomposition by $\sim 30\%$, with larger scatter at increased $f_{*}$. Our results are consistent with the findings in \citet{Yue_etal_2018} who decomposed SDSS-RM quasars into a central point source+host with ground-based deep imaging. During this comparison, we also investigated how different resolutions (e.g., seeing) and aperture sizes may impact the host-fraction measurements from ground-based imaging decomposition, using our HST images as the high-resolution counterparts. We found that typical seeing blurring and aperture effects (2\arcsec\ SDSS fibers) do not change our results. Therefore we conclude there are systematic differences in imaging and spectral decomposition to estimate the host starlight fraction. Nevertheless, this systematic difference in estimating host starlight contamination is not large enough to account for the systematic offset in the BLR radius-luminosity relation observed for the SDSS-RM sample \citep[][]{Grier_etal_2017,FonsecaAlvarez_etal_2019}.

Figure \ref{fig:fhost_m15} compares the host stellar masses derived from spectral decomposition in \cite{Matsuoka_etal_2015} and from imaging decomposition in this work. The spectral flux of host galaxies in \citet{Matsuoka_etal_2015} is corrected for fiber losses. Our stellar masses appear to be systematically smaller by $\sim$0.5\,dex, which might be due to different choices of initial mass functions (IMF) and simple stellar population (SSP) models: \cite{Matsuoka_etal_2015} used the \cite{Chabrier2003} IMF and the \cite{Maraston_etal_2011} SSP, while \cite{Into_Portinari_2013} and our {\tt CIGALE} fitting use the \cite{Kroupa2001} IMF and the \cite{Maraston2005} SSP.

\begin{figure*}
\centering
\includegraphics[width=0.8\textwidth]{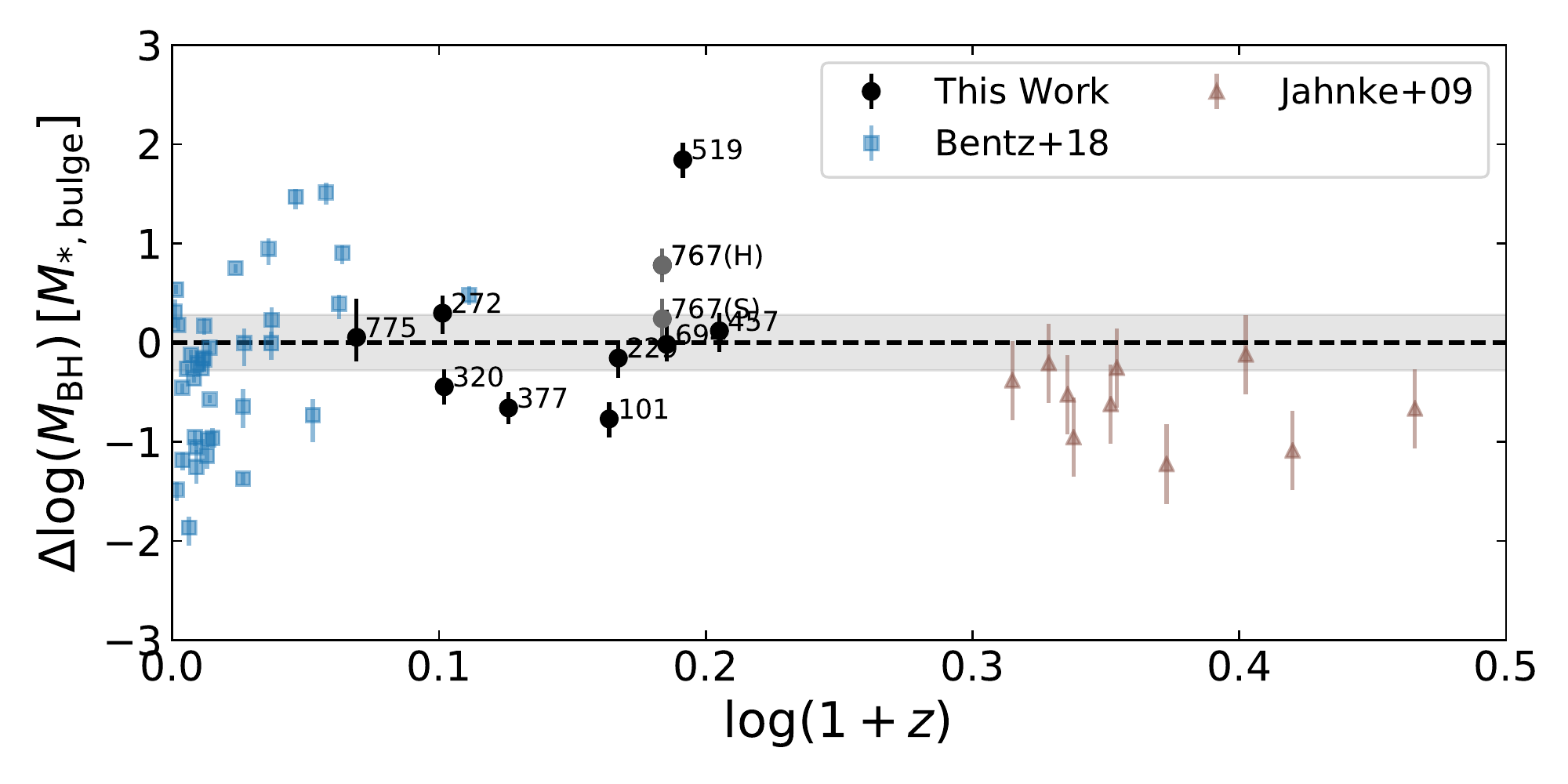}\\
\includegraphics[width=0.8\textwidth]{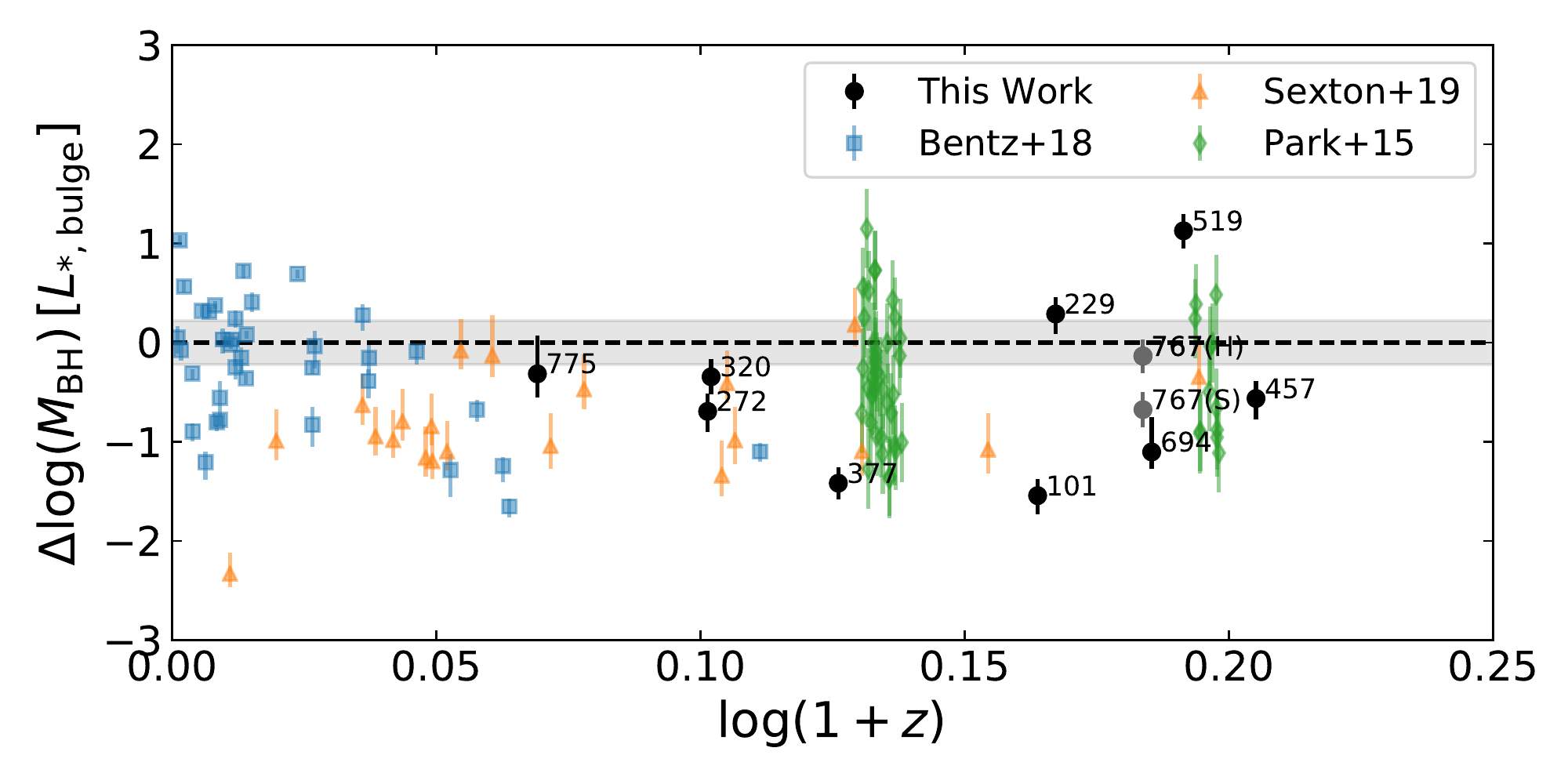}
\caption{Evolution of ${\rm \Delta log(M_{BH})}$ with redshift, with baselines adopted from the best-fit relations of ${\rm {M}_{BH}-{M}_{Bulge}}$ and ${\rm {M}_{BH}-{L}_{Bulge}}$ from the \cite{Kormendy_Ho_2013} sample. Vertical error bars are from uncertainties in BH mass only. } 
\label{fig:redshift}
\end{figure*}

\subsection{Redshift Evolution}\label{sec:evo}

The evolution of BH-host scaling relations with cosmic time is a key ingredient in understanding the origin of these correlations. As such, in recent years there have been numerous papers studying the cosmic evolution of the BH-host scaling relations \citep[e.g.,][]{Treu_etal_2004,Treu_etal_2007,McLure_etal_2006,Salviander_etal_2007,Jahnke_etal_2009,Bennert_etal_2010,Canalizo_etal_2012,Hiner_etal_2012,Salviander_Shields2013,Schramm_etal_2013,Busch_etal_2014,Park_etal_2015,Shen_etal_2015b,Matsuoka_etal_2015,Ding_etal_2017,Ding_etal_2020,Sexton_etal_2019}.

Figure \ref{fig:redshift} (upper panel) shows the deviation in $M_{\rm BH}-M_{\rm *,bulge}$ from the local baseline defined by quiescent galaxies \citep[ellipticals and classic bulges,][]{Kormendy_Ho_2013} as a function of redshift. $\Delta{\rm log_{10}}(M_{\rm BH})$ is consistent with zero within $<1.5\sigma$ for our sample (excluding the outlier RM519). Despite the large scatter compared to the local $M_{\rm BH}-M_{\rm *,bulge}$ relation (intrinsic scatter of $0.28$\,dex), there is no obvious evolution in the average deviation with redshift.

Figure \ref{fig:redshift} (lower panel) shows the deviation in $M_{\rm BH}-L_{\rm *,bulge}$ from the local baseline as a function of redshift. When $L_{\rm *,bulge}$ is not corrected for passive luminosity evolution (due to the aging of the stellar population), our sample, as well as the two other intermediate-redshift samples in \citet{Park_etal_2015} and \citet{Sexton_etal_2019} are consistent with the local $M_{\rm BH}-L_{\rm *,bulge}$ relation, albeit with larger scatter compared to that in the local baseline relation for quiescent bulge-dominant galaxies.

After correcting for passive luminosity evolution, \citet{Treu_etal_2007}, \citet{Bennert_etal_2010} and \citet{Park_etal_2015} reported evolution ($>3\sigma$ confidence level of evolution) in their sample (green diamonds in Figure \ref{fig:redshift}) for the $M_{\rm BH}-L_{\rm *,bulge}$ relation when compared to the local relation. However, our sample is consistent with the local $M_{\rm BH}-L_{\rm *,bulge}$ relation within $\sim2.5\sigma$ (excluding the outlier RM519) with no evolution in redshift when applying the same host luminosity correction \citep[equation 2 in][]{Park_etal_2015}.

Our sample covers by far the most extended redshift range with BH masses estimated directly from RM. Our uniform analysis of HST imaging decomposition does not reveal any noticeable evolution in the BH mass-bulge mass/luminosity relations over $0.2<z<0.6$. This is also consistent with the lack of evolution in the $M_{\rm BH}-\sigma_*$ relation measured for the SDSS-RM quasar sample \citep[][]{Shen_etal_2015b} over a similar redshift range. The sample size in this pilot study is small, and therefore we defer a more rigorous analysis of selection effects in constraining the evolution of the BH-host relations to future work.

\section{Conclusions}\label{sec:con}

Using high-resolution two-band HST imaging (UVIS and IR), we have measured the host and bulge stellar masses of ten quasars at $0.2\lesssim z\lesssim0.6$ with RM-based black-hole masses from the SDSS-RM project. Our quasars span more than one order of magnitude in BH and stellar masses. This represents the first statistical HST imaging study of quasar host galaxies at $z>0.3$ with direct RM-based black hole masses. 

We present the $M_{\rm BH}-M_{\rm *}$, $M_{\rm BH}-M_{\rm *,bulge}$ and $M_{\rm BH}-L_{\rm *,bulge}$ relations from our sample, and compare with local quiescent galaxies and other low-to-intermediate redshift AGN samples. Our quasars broadly follow the same BH--host scaling relations of local quiescent galaxies and local RM AGN. In addition, there is no significant evidence of evolution in the BH-host scaling relations with redshift. 

We compared our imaging decomposition with spectral decomposition in estimating the host starlight fraction. We found general consistency between the host fractions estimated with both methods. However, the host fraction derived from spectral decomposition is systematically smaller by $\sim 30\%$ than that from imaging decomposition, consistent with the findings using ground-based imaging \citep{Yue_etal_2018}. 

While the sample size in this pilot study is too small to provide rigorous constraints on the potential evolution of the BH-host scaling relations and assess the impact of selection effects, it demonstrates the feasibility of our approach. We are acquiring HST imaging for 28 additional SDSS-RM quasars at $0.2<z<0.8$ with direct RM-based BH masses, which will enable more stringent constraints on the evolution of the BH-bulge scaling relations up to $z\sim 1$.

\acknowledgements

We thank C.~Y.~Peng for useful suggestions on GALFIT, and K. A. Phadke for useful discussions on CIGALE. YS acknowledges support from an Alfred P. Sloan Research Fellowship and NSF grant AST-1715579. Support for this work was provided by NASA through grant number HST-GO-14109 from the Space Telescope Science Institute, which is operated by AURA, Inc., under NASA contract NAS 5-26555. LCH was supported by the National Key R\&D Program of China (2016YFA0400702) and the  National Science Foundation of China (11721303, 11991052). EDB is supported by Padua University
grants DOR1715817/17, DOR1885254/18, and DOR1935272/19 and by
MIUR grant PRIN 2017 20173ML3WW\_001.

\software{
AstroDrizzle, 
Astropy \citep{astropy:2013,astropy:2018}, 
CIGALE \citep{CIGALE}, 
GALFIT \citep{galfit}, 
matplotlib \citep{matplotlib}, 
Numpy \citep{numpy}, 
{\tt photutils} \citep{photutils}, 
{\tt pysynphot} \citep{pysynphot}, 
seaborn \citep{seaborn}.}

\facility{HST (WFC3/UVIS, WFC3/IR)}

\bibliography{refs}

\end{document}